\documentclass[reprint,aps,amssymb,amsmath,superscriptaddress]{revtex4-1}

\usepackage{graphicx}
\usepackage{dcolumn}
\usepackage{fancyref}
\usepackage{bm}
\usepackage{color}
\usepackage[toc,page]{appendix}
\usepackage[colorlinks=true,allcolors=blue]{hyperref}
\usepackage[normalem]{ulem}
\usepackage[fleqn]{mathtools}
\usepackage{etoolbox}

\newcommand{\vvec}[1]{\mathbf{#1}}
\newcommand{\ttens}[1]{\mathrm{\mathbf{#1}}}

\def\dee{\mathrm{d}} 
\def\cc{c}

\begin{document}

\title{Conformal Elasticity of Mechanism-Based Metamaterials}

\begin{abstract}
    
    Deformations of conventional solids are described via elasticity, a classical field theory whose form is constrained by translational and rotational symmetries. However, flexible metamaterials often contain an additional approximate symmetry due to the presence of a designer soft strain pathway. Here we show that low energy deformations of designer dilational metamaterials will be governed by a novel field theory, conformal elasticity, in which the nonuniform, nonlinear deformations observed under generic loads correspond with the well-studied---conformal---maps. We validate this approach using experiments and finite element simulations and further show that such systems obey a holographic bulk-boundary principle, which enables an analytic method to predict and control nonuniform, nonlinear deformations. This work both presents a novel method of precise deformation control and demonstrates a general principle in which mechanisms can generate special classes of soft deformations.

\end{abstract}


\author{Michael Czajkowski}
\affiliation{School of Physics, Georgia Institute of Technology, Atlanta, Georgia 30332, USA}
\author{Corentin Coulais}
\affiliation{Institute of Physics, Universiteit van Amsterdam,  Science Park 904, 1098 XH Amsterdam, The Netherlands}
\author{Martin van Hecke}
\affiliation{AMOLF, Science Park 104, 1098 XG Amsterdam, The Netherlands }
\affiliation{Huygens-Kamerlingh Onnes Lab, Universiteit Leiden, PObox 9504, 2300 RA Leiden, The Netherlands}
\author{D. Zeb Rocklin*}
\affiliation{School of Physics, Georgia Institute of Technology, Atlanta, Georgia 30332, USA}
\date{\today}

\maketitle


\section*{Introduction}

\begin{figure*}[t]  
\begin{center}
    \includegraphics[width=0.8\textwidth]{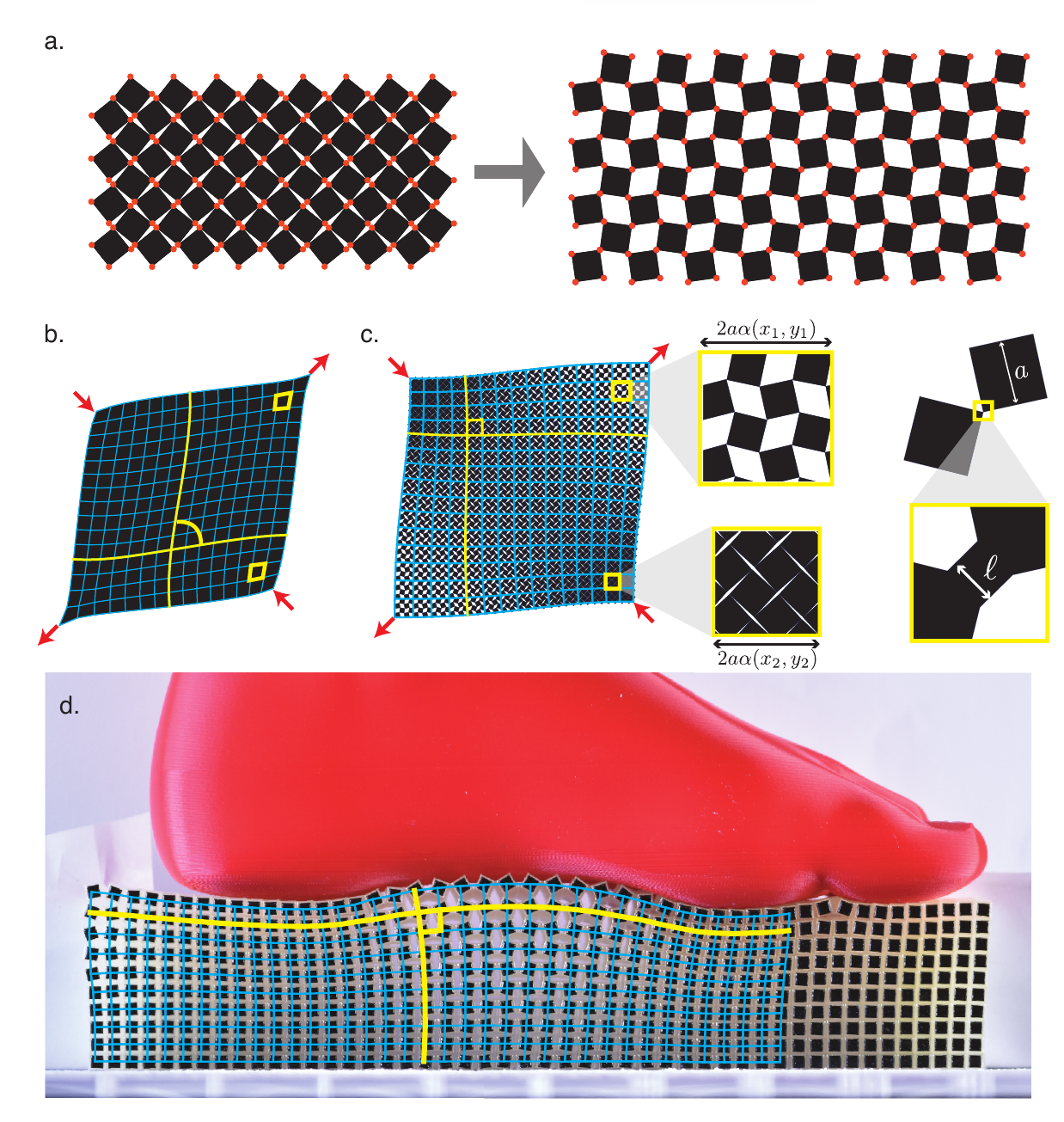}
    \caption{ 
    \textbf{Dilational Mechanism and Conformal Deformations.}
    (\textbf{a}) In the ideal rotating square mechanism, a structure of rigid squares (black) connected by frictionless hinges (red), may be dilated and contracted at zero energy cost when the squares are rotated opposite to their neighbors.
    (\textbf{b}) Applying fixed deformations (red arrows) at the boundary of a conventional elastic material leads to a spatially varying strain field that includes shear components that change the local angles between pieces of material, such as the initially perpendicular grid lines (blue, yellow).
    (\textbf{c}) In contrast, a pure dilational metamaterial designed around the RS mechanism, may accommodate the loading without shearing of the unit cells, so that even as grid lines rotate, the angles between them remain fixed.
    This angle-preserving behavior arises due to a local version of the RS mechanism, in which square elastic chunks of side length $a$ rotate about small, flexible hinges of thickness $\ell$ to open and close pores according to $\alpha(x,y)$, the local linear dilation factor. (\textbf{d}), A fabricated sample of checkerboard material approximately preserves right angles (shown in yellow) under a generic nonlinear ``foot'' load, suggesting conformal deformation behavior. 
    \label{fig:checkerboard_diag} }
    \end{center}
\end{figure*}

Mechanical metamaterials use patterns of
cuts, pores and folds to achieve nonlinear~\cite{MIURA1985, Bertoldi2010, Shim2012}, programmable~\cite{Florijn2014, Silverberg2014}, polar~\cite{Kane2014, Paulose2015, Saremi2020} and other exotic behavior~\cite{Milton1995, Filipov2015, Deng2019, Deng2020, Jin2020, Bertoldi2017} in response to external forcing. Often these features rely on the careful arrangement of the cuts, pores and folds to emulate a mechanism, which is a special pathway of deformation that enables the metamaterial to change shape at very small (ideally zero) energy cost. For example, the canonical rotating square (RS) mechanism, which consists of perfectly hinged rigid squares, enables a uniform dilational motion (Fig.~\ref{fig:checkerboard_diag}a) and has inspired the design of a range of metamaterials which collapse inward rather than bulge outward when compressed from the lateral direction~\cite{Grima2000, Mullin2007, Bertoldi2010}.  While the dilational mechanism becomes a true zero energy motion in the limit of vanishing hinge size, for realistic metamaterials with finite hinges this uniform dilational motion is only observed for the particular case of a completely homogeneous loading condition. For generic, i.e. inhomogenous, loading conditions the response is more complex~\cite{Coulais2018}, and yet a general framework to describe the general nonuniform soft deformations of mechanism-based metamaterials is lacking. 

Here, we aim to decode the nonuniform deformations of metamaterials based on a dilational mechanism. While elastic response of ordinary materials will locally be composed of some finite portion of shear (Fig.~\ref{fig:checkerboard_diag}b), a dilational material will strive to expel shear everywhere in favor of the dramatically softer dilational strains. Hence, one may wonder whether the nonuniform deformations of dilational metamaterials may be locally composed of pure dilation with no shear? 
While not all spatial patterns of
strain are realizable due to basic geometric restrictions from compatibility relations, conformal maps are well-known to constitute the full set of compatible smooth deformations which locally are composed of dilational strains only.
These maps therefore provide a recipe to compose nonuniform soft modes from slow spatial variation of e.g. the RS mechanism (Fig.~\ref{fig:checkerboard_diag}c). We therefore formulate our conformal hypothesis: under a generic and broad set of loading conditions, dilational metamaterials will respond with an angle-preserving conformal deformation (Fig.~\ref{fig:checkerboard_diag}d), energetically much softer than conventional elastic strains. 

In this work, we confirm this hypothesis using simulations, experiments, and a coarse grained elastic theory. We first show that the response of the RS metamaterial is indeed conformal at both global and local scales. We then present a reduced elasticity theory which facilitates analytic insight, including new methods of predicting nonlinear deformation. We then use the bulk-boundary correspondence principle obeyed by these metamaterials to introduce a recipe for on-demand activation of each soft configuration.

\section*{Results}
\begin{figure*}[!t]  
\begin{center}
    \includegraphics[width=0.9\textwidth]{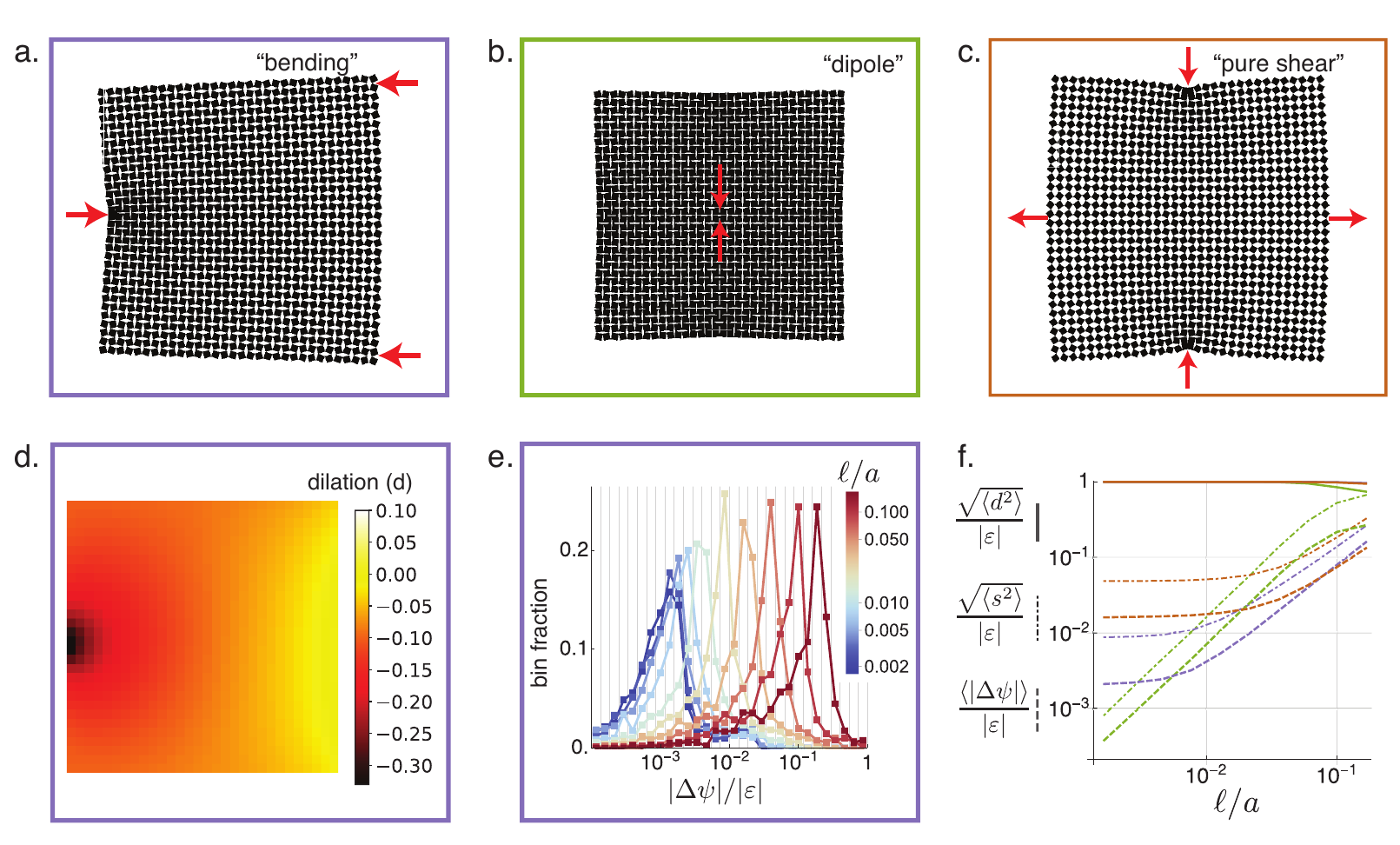}
\caption{\label{fig:op_simulations}   \textbf{Deformations of dilational metamaterials are angle preserving.} 
  (\textbf{a}) A Finite-Element Simulation of a three-point bending test (the ``bending'' load) locates force balanced (relaxed) states of the RS-based elastic metamaterial 
  subject to displacements on the boundaries that are incompatible with uniform dilation.
    (\textbf{b}) The ``dipole'' load consists of closing a single central pore by displacement of points on opposite sides of the pore, thereby generating a localized dilation in the bulk of the material.
    (\textbf{c}) The ``pure shear'' load consists of compression along one axis and extension along the other, which is nevertheless compatible with a spatially varying pure dilation field.
    (\textbf{d}) The resulting dilation field under the ``bending'' load varies widely across the sample, including nonlinear compression and expansion. 
    (\textbf{e}) Nevertheless, the angle changes $|\Delta \psi|$ remain small, even in proportion to the average strain magnitude $|\epsilon|$. Here, the histogram is shown on log-linear scale, with each square marker giving the fraction of unit cells with that angle, which tend to decrease as the hinge size decreases. The preservation of angles is the defining feature of conformal maps.
    (\textbf{f}) For all loads (colored in correspondence with the box outlines in (\textbf{a}), (\textbf{b}), (\textbf{c})) the dilation fraction (solid lines) is nearly unity, and the deviation from unity is captured by the shear fraction (dot dashed lines), which for linearly small shears is the complement of the dilation fraction, and at the smallest hinge size ranges between $0.001 - 0.05$. Both the shear fraction and the normalized average angle change (dashed lines) decrease with decreasing hinge size. The plateau in the bending and pure shear loads can be attributed to non-continuum lattice effects.  
}
    \end{center}
\end{figure*}

To test the hypothesis of conformal deformation, we investigate the elastic response of RS-based metamaterials at a range of hinge thicknesses using 
finite element simulations which preserve the intricate pore structure (see Methods).
While the detailed microscopic strains may contain significant shear, we use the displacement of the square centers to extract coarse-grained shear strain and dilational strain fields, with the expectation that the latter should dominate as the hinges become small. 
Even under nonuniform strain, dilation indeed dominates over shear in simulations of three-point ``bending'' (Fig~\ref{fig:op_simulations}a), a local ``dipole'' dilation (Fig~\ref{fig:op_simulations}b) and even when the system is subject to global ``pure shear'' via compression along one axis and expansion along the other (Fig~\ref{fig:op_simulations}c). Even for dilations that vary dramatically through space (Fig.~\ref{fig:op_simulations}d), the average amount of local shear compared to dilation is captured in the shear fraction, defined in Appendix 1, and varies from $10^{-1.5}$ to $10^{-3}$, in contrast with  values on the order of one in conventional structures (Fig~\ref{fig:op_simulations}f).
The low amounts of shear also imply nearly preserved angles, which is the  defining characteristic of conformal deformations.
As confirmed in Fig.~\ref{fig:op_simulations}e,f, the average angle change and the shear fraction both decrease as the hinge thickness is reduced and the system approaches the ideal mechanism limit.
Despite finite-size effects from the $16\times16$ unit cell lattice, the strong preservation of angles indicates that the response is locally conformal. 

\subsection*{Fitting Deformations with Conformal Maps}

\begin{figure*}[!t]  
\begin{center}
    \includegraphics[width=0.9\textwidth]{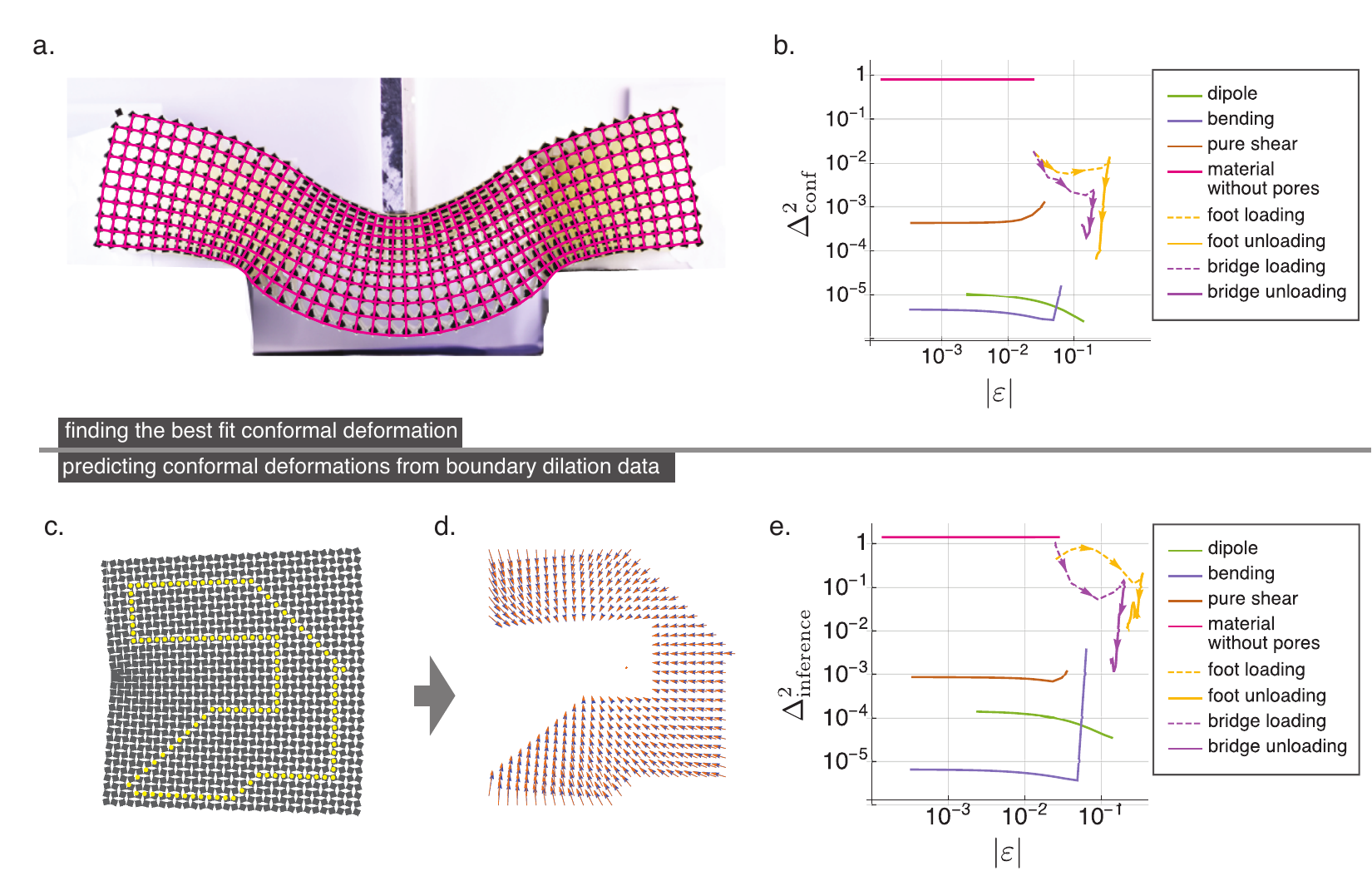}
    \caption{\label{fig:boundary_method} 
\textbf{Deformations are conformal and permit nonlinear prediction.}
(\textbf{a}) The nonlinear deformation of a three-point ``bridge'' load can be well fit to a conformal map (magenta grid) chosen to minimize the average squared deviation between the mapped and observed displacements.
(\textbf{b}) The fraction of variance unexplained $\Delta^2_{\textrm{conf}}$ between observed deformations and conformal fits in both simulations and experiments is very small, demonstrating that the soft deformations correspond globally to conformal maps as well as locally to pure dilations. Experimentally, the fit is especially good in the unloading step, in which stresses arising from frictional interactions at the material boundary are able to relax.
In contrast, the data for a conventional elastic material (pink) cannot be described as a conformal map.
(\textbf{c}) According to the holographic property of conformal maps, the dilation field on the boundary of an arbitrarily chosen (gerrymandered) region, such as the one depicted in yellow for a simulation configuration, should uniquely determine the deformation within the region.
(\textbf{d}) Indeed, the displacements inferred from the boundary dilations (orange arrows) match closely with those observed directly in simulations (blue arrows).
(\textbf{e}) The observed displacements are now compared to the conformal maps inferred in this way, and the fraction of variance unexplained plotted as $\Delta^2_{\textrm{inference}}$. This demonstrates that the deformations are not merely conformal, but the precise conformal map is encoded on the boundary. Here, the predictions of experimental deformations are inferred from the full rectangular boundary, while predictions of FEM data are all inferred from the gerrymandered boundary points shown in (\textbf{c}) to show robustness to the choice of boundary. Note that the example chosen for display in \textbf{(c,d)} corresponds to the largest strain explored in the bending load, and according to \textbf{(e)} has the largest prediction error of the FEM data, and all other FEM data is predicted to even greater accuracy than depicted here.
 }
    \end{center}
\end{figure*}

The observation of approximately conformal local behavior 
suggests that deformations of the elastic RS metamaterial may also correspond globally to a conformal map. This motivates an elegant formulation expressing positions in the plane as complex numbers $\mathbf{r} = (x,y) \leftrightarrow x+iy \equiv z$ which has previously been applied to elasticity in a variety of contexts~\cite{england2003complex}. In our case, it has the added utility that any exact conformal deformation takes the convenient form
\begin{equation}\label{eq:conformalmap}
f(z) = \sum_{n=0}^{\infty}C_n z^n,
\end{equation}
where  $f(z)$ is the deformed position of the material element initially located at $z$ and the complex coefficients $\{C_n\}$ define the map. 
The ability to express such a map in this reduced series form, turns the search for the conformal map closest to our data into a linear algebra problem which is readily solved. Using only the first 20 $C_n$ coefficients, we use this procedure in Fig.~\ref{fig:boundary_method}a,b to identify conformal maps that are able to capture all but a small fraction $\Delta_{\textrm{conf}}$ (about $1\%$) of the observed displacements. 
Note here that, to appropriately assess the error in many different approximations of the observed square block displacements, we employ the fraction of variance unexplained $\Delta^2$. This measure is equivalent to the mean square displacement error normalized by the mean square magnitude of displacements; an explicit formula for this is given in Appendix 2. Indeed, despite the nonzero local angle deviations, we find that RS-based metamaterials will respond to loading with a deformation that is closely matched with a conformal map even under nonlinear strains. This is in contrast with generic deformations, which fit poorly to equation~(\ref{eq:conformalmap}) as shown in Fig.~\ref{fig:boundary_method}b and in general require terms with complex conjugation of $z$ such as $z^2 \bar{z}^5$ included in the expansion to be captured accurately.



\subsection*{Conformal Elasticity}

Given this evidence that the RS metamaterial responds to loading with a near-conformal deformation, we present an elastic energy functional,
\begin{align}\label{eq:fullenergy}
E =\int d^2 \mathbf{r}  \frac{1}{2} \Big[ G_1(\alpha) s_1^2 + G_2(\alpha) s_2^2 + \qquad \qquad \qquad
\\ \nonumber 
\qquad \qquad \qquad
\frac{\ell^2}{a^2} M(\alpha) + a^2 \tilde{M}(\alpha) |\nabla{\alpha}|^2\Big] \,
\end{align}
which the system minimizes subject to the boundary conditions. Here, $\alpha$ is a (nonlinear) spatially varying field describing the dilation factor of the structure relative to its equilibrium, while $s_1$ and $s_2$ are the (linear) coarse pure shear and simple shear respectively, $\ell$ is the width of a hinge and $a$ the width of a square piece. $G_{1,2}(\alpha)$ are shear moduli, $M(\alpha)$ the dilation energy density and $\tilde{M}(\alpha)$ the modulus associated with spatial variations in the dilation. These fields may be defined more precisely in terms of the right Cauchy-Green deformation tensor $\ttens{C}$ (the metric of deformation). When we choose to orient the axes of our reference space along the vectors connecting square centers to those of their neighbors they take the form
 \begin{align}\label{eq:define_deformation}
 \alpha & = (\mathrm{det}[\ttens{C}])^{1/4}, \\
s_1 & = \frac{\mathrm{Tr}[\ttens{C}\cdot \ttens{\sigma}^{(3)}]}{\sqrt{\mathrm{det}[\ttens{C}]}},  \\
s_2 & = \frac{\mathrm{Tr}[\ttens{C}\cdot \ttens{\sigma}^{(1)}]}{\sqrt{\mathrm{det}[\ttens{C}]}} \, ,
\end{align}
where $\ttens{\sigma}^{(1)}$ and $\ttens{\sigma}^{(3)}$ are the first and third Pauli matrices.
While alternate analytic approaches to the nonlinear strain of anisotropic media have been established~(e.g.,\cite{Ericksen1997}), this formalism is particularly convenient for the RS metamaterial,
for which any additional terms such as couplings between dilation and shear are excluded by scaling arguments and by the the orthotropic symmetry (p4g) as presented in Appendix 5.
 This specific form in equation~\ref{eq:fullenergy} may also be derived by a nonlinear coarse graining, shown in Appendix 4, which provides additional useful insight into the moduli. 
 Although other continuum elasticity theories for the RS metamaterial have been developed in one~\cite{Coulais2018} and two dimensions~\cite{Deng2019, Deng2020}, and another theory for a bistable dilational material~\cite{Jin2020}, this nonlinear analytic form, which centers the dilational mechanism and its gradients, has not been proposed previously. While this energy function holds a rare level of insight into nonlinear deformation, it is still quite difficult to solve analytically.

 To gain more analytical traction, we employ perturbation theory in the limit in which the hinges are tiny relative to the unit cell size $l/a \ll 1$ (i.e. the soft mechanism limit) and the material sample is composed of very many unit cells (i.e., the continuum limit). Here, the second two terms in equation~\ref{eq:fullenergy} may be regarded as small perturbations, and the energy cost of deformations that include shear (first two terms of equation~(\ref{eq:fullenergy})) become prohibitively stiff by comparison. Therefore, in this limit, dilations will begin to act as a local symmetry and the conformal maps will constitute a degenerate space of ground states of the energy. Restricting our focus to this space of ground states, we recover a nonlinear notion of conformally invariant elastic theory, which was suggested by Sun. \textit{et al.}~\cite{Sun2012} previously for the linear deformations of the Kagome lattice, but has otherwise been viewed as ``unphysical''~\cite{Riva2005} until now. 
 
 With the nonperturbative part of the energy penalizing only shear, and the set of conformal maps forming a degenerate set of ground states, it would seem that an infinite number of conformal maps are equally likely to arise in response to generic loading such as that in the FEM simulations and experiments of Fig.~\ref{fig:op_simulations}. However, this is not the case, as the degeneracy is broken by the perturbative part of the energy functional
\begin{equation}\label{eq:conformal_elasticity}
\begin{aligned}
\Delta E \equiv \int d^2 z \frac{1}{2} \left[ \frac{\ell^2}{a^2}M(|f'(z)|)+ a^2 \tilde{M}(|f'(z)|) |f''(z)|^2\right],
\end{aligned}
\end{equation}
 which is now expressed in terms of the conformal map $f:z \rightarrow f(z)$ describing the local dilation and rotation, with $f'(z) \equiv \alpha e^{i \phi}$ (equation~(\ref{eq:conformalmap})). 
 While the usual problem of reducing to a constrained elasticity theory is typically done using Lagrange multipliers, conformal maps simply have $s_1 = s_2 = 0$ and $\alpha = |f'|$, allowing equation~(\ref{eq:conformal_elasticity}) to be obtained easily as the conformal limit of equation~(\ref{eq:fullenergy}). The alternate route, utilizing Lagrange multipliers, is explored in Appendix 8, and yields useful information about stress. 
The energy in equation~(\ref{eq:conformal_elasticity}) arises purely from the last two terms in equation~(\ref{eq:fullenergy}) and simultaneously breaks the conformal invariance and the ground state degeneracy, allowing predictions of specific conformal response to be generated; we refer to this procedure as ``conformal elasticity''.  
With this, we have reduced the difficult tensorial problem of conventional nonlinear elasticity of a material with pores down to a scalar theory which is much more analytically tractable.
For small loading, minimizing the energy and thus predicting the deformation is reduced to a linear algebra problem, which is readily solved. As shown in Appendix 7, these predictions closely match the observed finite-element displacements with correlation coefficient $R^2 \approx 0.99 - 0.9999$ for hinges with $\ell/a = 0.005$. This result showcases both the accuracy of conformal elasticity and the mathematical convenience of conformal maps.

\subsection*{Bulk-Boundary Correspondence Generates Accurate Nonlinear Analytic Predictions}

Solving equation~(\ref{eq:conformal_elasticity}) analytically is only possible in the linear regime and relies on prior knowledge of the effective stiffnesses $M$ and $\tilde{M}$. However, as shown in Fig.~\ref{fig:boundary_method}a,b, the conformal property itself extends to nonlinear deformations. We therefore devise a scheme to extend analytic predictions to the nonlinear regime, relying on a mathematical property of conformal maps: bulk-boundary correspondence. Within our formalism, this bulk-boundary correspondence works as follows: if we know the amount of dilation along the entire boundary of a section of RS material, we can predict analytically the local dilation everywhere in the material interior from the unique conformal map derivative $f'(z)$ consistent with the boundary conditions and can further integrate to infer the displacements. We illustrate the principle in Fig.~\ref{fig:boundary_method}c,d,e. We retrieve the dilation on an arbitrarily chosen yellow domain in Fig.~\ref{fig:boundary_method}c from the FEM data, use the bulk-boundary correspondence to calculate the deformation inside this domain (See displacement field depicted by blue arrows in Fig.~\ref{fig:boundary_method}d), and show that these are in excellent agreement with the observed displacement data (See displacement field depicted by orange arrows in Fig.~\ref{fig:boundary_method}d). This method is robust to the arbitrary choice of boundary shape and captures $>99\%$ of block displacements even at large strains (Fig~\ref{fig:boundary_method}e).

\subsection*{Bulk-Boundary Correspondence Generates a Method for Precise Deformation Control}

The bulk-boundary correspondence can also be used for prescribing on-demand deformations. Given a target deformation field of some section of material, we can determine the suitable actuation pattern by simply examining the local target dilation  at the material boundary. We illustrate this with three different actuation patterns in Fig.~\ref{fig:conformal_fun}a,b,c. Within a simulation described in Appendix 10, we actuate the material at its boundary (the red bars represent actuators) and observe that the numerically force balanced deformation is in good agreement with the yellow dots that mark the target centers of the squares. Of course in addition to being conformal, this procedure is only able to actuate deformations that do not stretch or compress the material beyond the physical bounds of the mechanism itself. This task is aided by the maximum modulus principle of conformal maps, in which both the maximum and minimum dilations must occur on the material boundary. We can therefore guarantee that for any choice of actuation that varies slowly along the boundary, there is a physically valid soft conformal deformation that will be activated in the interior. We propose that such boundary-control may be achieved via the insertion of struts in the pores near the boundary, which, with remote control actuation, would allow these soft modes to be activated remotely on-demand. Exploration of this possibility, with applications in development of soft robotics, is reserved for future work.

While the deformations of Fig.~\ref{fig:conformal_fun} arise from relatively simple functions for the input boundary dilation patterns, each analytic form of the interior deformation turns out to be much more complicated. However, one limit is simple enough to offer intuitive insight. When the lengthscales at which the dilation varies are much shorter than those at which the boundary curves, the boundary may be treated as that of a semi-infinite plane. For conformal metamaterial occupying the infinite upper half plane ($y>0$), and the dilations along the boundary (x-axis) constrained to satisfy $\alpha(x, y=0) = \alpha_0(x)$, there is again only one allowed pattern of dilation, which takes the form
\begin{equation}
        \alpha(x,y) = \text{exp}\left[ \int_{-\infty}^\infty \mathrm{d}k e^{ikx - |k|y} w(k) \right] \, ,
\end{equation}
where $w(k)$ is the Fourier transform of the function $w(x) = \text{ln}(\alpha_0(x))$. In this form, we can see that the dilation (or, rather, the logarithm of the dilation) decays into the bulk exponentially according to the lengthscale of variation of the boundary dilation. For instance, when $w(x)= w_0 \text{cos}(k_0 x)$, we will have $\alpha(x,y) = \text{exp}\left[ w_0 e^{-k_0 y} \text{cos}(k_0 x)  \right]$.

\begin{figure*}[!t]  
\begin{center}
    \includegraphics[width=0.85\textwidth]{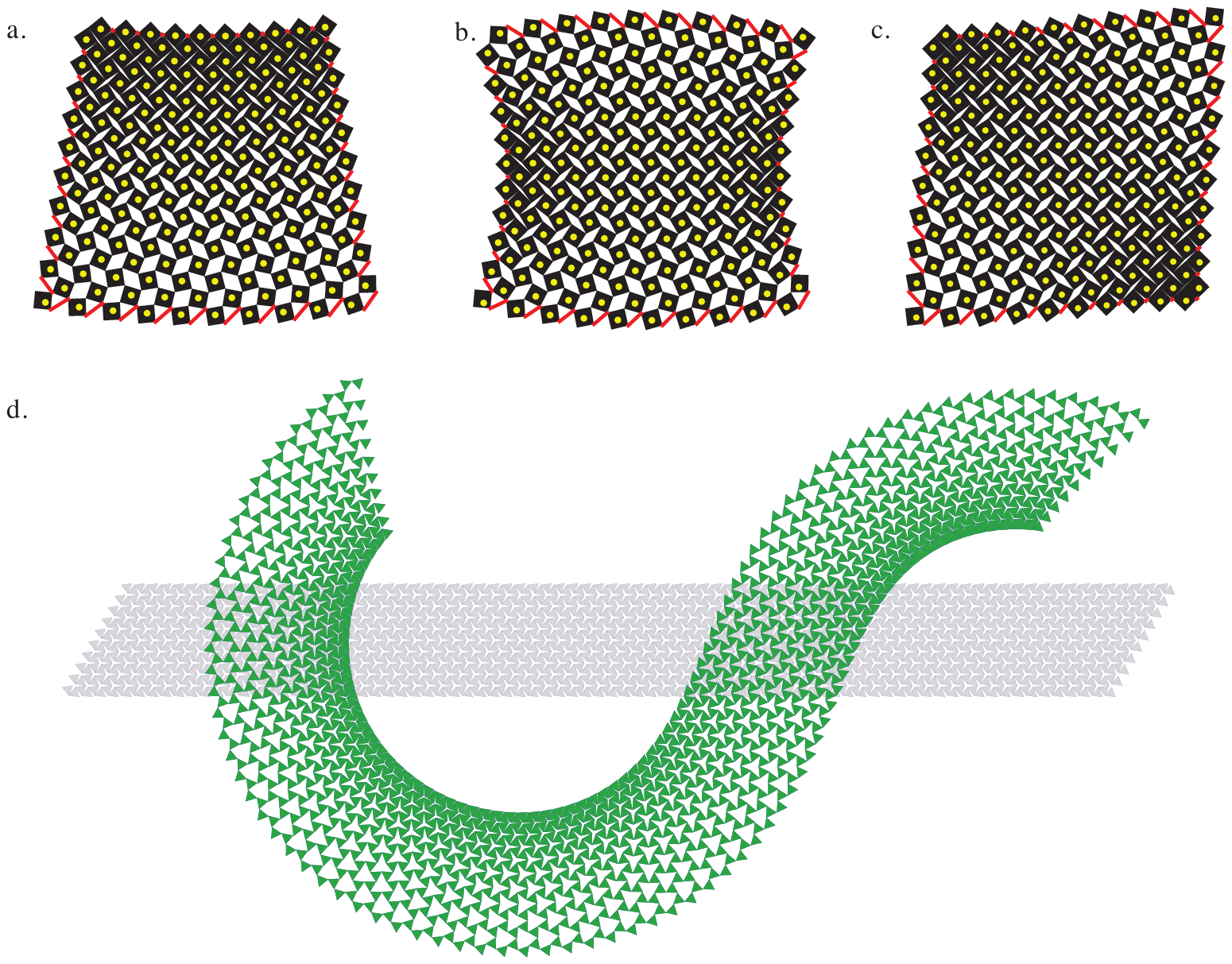}
    \caption{\label{fig:conformal_fun} \textbf{Bulk-Boundary correspondence allows for precise control of metamaterial deformation.}
    (\textbf{a, b, c}) The bulk-boundary correspondence shown in Fig.~\ref{fig:boundary_method} enables
    on-demand actuation of material deformation via the addition of boundary springs (red). Here, the locations of square centers observed in simulation match well with analytically obtained target positions (yellow dots) for force-balanced configurations of a simplified ball-and-spring model.
    (\textbf{d}) Because the principles presented here extend to any dilational metamaterial, the kagome lattice may likewise be placed in a low-energy conformal deformation. The kagome lattice is capable of undergoing a broader range of deformations without self-intersection, permitting a greater range of controllable deformations such as pictured here.
}
    \end{center}
\end{figure*}

\section*{Discussion}

The intuitive concept of soft modes that are locally composed of a spatially varying mechanism is not confined to the RS mechanism explored here. The conformal modes explored here arise because they are the Nambu-Goldstone modes associated with the local dilational symmetry, rather than from the details of the specific microstructure. We therefore propose the analogy: as isometries are to thin elastic sheets~\cite{Santangelo2013}, so are conformal deformations to dilational metamaterials.
Consequently, dilational materials derived from a variety of different metastructures~\cite{Prall1997, Sun2012, Milton2013, Cho2014, Gatt2015} including fractal RS structures~\cite{Cho2014, Gatt2015}, disordered ``pruned'' networks~\cite{Goodrich2015}, nanoscale~\cite{Suzuki2016}, and three dimensional generalizations~\cite{Shim2012,Konakovic2016, Choi2019} will similarly have mechanical response controlled by a set of conformal soft modes. These alternate architectures provide the possibility for greater ranges of dilation~\cite{Milton2013}, opening the door for even more dramatic soft deformations such as pictured for the kagome lattice in Fig~\ref{fig:conformal_fun}d. 

Note, by example, that many possible loads applied to thin sheets (e.g. applying an in-plane stretch to a flat sheet) are incompatible with isometries, and an isometric soft mode theory must break down there. Similarly, a variety of overly strict loads will be incompatible with conformal deformation and our conformal elasticity theory must similarly break down. For instance, this will happen when attempting to constrain the displacements, rather than dilations all along a closed boundary. Conveniently, we have identified two particular scenarios with either a finite number of sufficiently spaced point displacements, or with dilations controlled along a closed boundary, each of which contain a broad variety of loading possibilities which are guaranteed to be compatible with a conformal map, and are therefore governed by our elasticity theory. Identifying theories that apply beyond the soft conformal response will provide an interesting avenue for future work.

We suggest that a broad class of generic mechanisms not confined to pure dilation will also generate families of soft modes that govern material response, as were indeed observed in so-called ``kirigami'' structures~\cite{Zheng2021} and that this may become a fundamental principle for mechanism-based metamaterials, with potential applications from footwear to soft robotics.
In addition, topologically polarized systems~\cite{Kane2014, Rocklin2017}, which have additional mathematical structure controlling an additional set of exotic boundary modes, necessarily break the symmetry of the mechanism strain pathway that allows conformal modes and the notion of mechanism compatibility may be extended to their nondilational mechanisms while incorporating topological notions.
Exploring how these new classes of soft modes may obey, e.g., a generalized bulk-boundary correspondence may yield fruitful connections to exotic field theories~\cite{Pretko2018} as well as black hole and string theories where holographic principles already play a vital role.

\section*{Methods}
\label{sec:methods}

\subsection*{Generating deformations analytically}

We analytically generate the deformations of two-dimensional metamaterials that locally resemble the uniform mechanism and a rotation and hence eliminate the dominant contributions to the elastic energy, taking advantage of a map between the plane of deformation and the complex plane. In general, a deformation  map $\vvec{x} \rightarrow \vvec{X}(\vvec{x})$ has deformation tensor  $\ttens{F}_{ij} \equiv \frac{\partial X_i }{\partial x_j}$ which deforms infinitesimal material vectors as $\dee \vvec{x}_i \rightarrow \ttens{F}_{ij}  \dee \vvec{x}_j$. To correspond locally to a rotation of the mechanism field, the deformation tensor must be proportionate to a rotation matrix. We seek solutions in which the rotation angle $\phi$ and the dilation factor $\alpha$ vary over space, so that $\ttens{F} = \alpha \ttens{R}(\phi)$. However, this generates nontrivial restrictions because of the requirement that the deformed state not tear, i.e. that $\oint \dee \mathbf{X} = 0$ along any path. By applying Stokes' theorem, this leads to the standard geometric or kinematic compatibility conditions:

\begin{align}
    \partial_1 F_{12} - \partial_2 F_{11} &= 0, \\
        \partial_1 F_{22} - \partial_2 F_{21} &= 0.
\end{align}

This generates two independent conditions on the two fields, suggesting a unique solution subject to certain boundary conditions. However, it becomes convenient to define new variables $z=x+iy, \bar{z} = x - i y$, with $i$ the imaginary unit. Enforcing the conditions $\partial_z z = \partial_{\bar{z}}\bar{z} = 1, \partial_z \bar{z} = \partial_{\bar{z}} z = 0$ then requires that $\partial_z = (1/2)(\partial_x - i \partial_y)$, $\partial_{\bar{z}} = (1/2)(\partial_x + i \partial_y)$. 

By summing the second compatibility condition with $(-i)$ times the first, we find, following algebra, that compatibility is equivalent to the complex condition $\partial_{\bar{z}} \left[\alpha \exp(i \phi)\right] = 0$. Thus, the locally dilational deformations of the metamaterial are exactly the conformal maps of the plane, which may be expressed as analytic functions $f'(z)$ whose magnitude is the local dilation and whose argument is the local orientation. Displacements can be obtained via complex integration: $f(z) = \int \dee z f'(z)$. This method hints at a more general class of spatially varying deformations in unimode materials, but the results in this case follow from the fact that the shear-free deformations of the structure are precisely the angle-preserving ones well-known to be described by complex analytic functions.


\subsection*{Finite Element Simulation Protocol \label{subsec:fem_protocol}}

For the 2D finite-elements simulations, we use the commercial software Abaqus/Standard. We use a neo-Hookean energy density as a material model, a shear modulus $\mu =0.333$ MPa*m, bulk modulus
$K_0 = 16.7$ GPa*m and plane strain conditions with hybrid quadratic triangular elements (Abaqus type CPE6H). 
We construct the mesh so that the thinnest parts of the samples are at least two elements across. 
We perform two types of simulations (i) on the metamaterials unit cells with periodic boundary conditions, on which we perform a low strain static test ($5\times10^{-6})$) and (ii) on the full structure, on which we perform static nonlinear analysis. 

In the simulations (i), to implement periodic boundary conditions, we define constraints on the displacements of all of the nodes at the horizontal and vertical boundaries of the unit cell and impose displacement using virtual nodes~\cite{Coulais2016}. We perform four types of simulations to extract the macroscopic linear bulk modulus $K = \frac{\ell^2 M''(\alpha=1)}{4 a^2}$, and shear moduli $G_1(\alpha=1), G_2(\alpha=1)$. In order to extract $K$ and $G_1$, we apply biaxial compression $\varepsilon^\text{comp}_{11}=\varepsilon^\text{comp}_{22}=5\times 10^{-6}$, $\varepsilon_{12}=0$ and pure shear $\varepsilon^\text{shear}_{11}=-\varepsilon^\text{shear}_{22}=5\times 10^{-6}$, $\varepsilon_{12}=0$ to a unit cell with periodic boundary conditions. $K$ and $G_1$ are each extracted as a slope in linear fitting  $\sigma^\text{comp}_{11} = 2 K \varepsilon^\text{comp}_{11}$ and $\sigma^\text{shear}_{11} = 2 G_1 \varepsilon^\text{shear}_{11}$. $G_2$ is extracted using the identical procedure as for $G_1$ applied to a unit cell that is rotated by $\pi/4$. We examine a unit cell with $a=12$mm, pretwist $\theta_0 = 0.39$ and hinge thickness $\ell=0.1$mm, finding $(K, G_1, G_2) \sim (35, 5.2*10^4, 2*10^4)$Pa*m. These modulus measurements are essential for obtaining the perturbative energy minimizing prediction of deformation, summarized briefly in the main text and given in more detail in Appendix 7.

In the simulations (ii), we use three kinds of boundary conditions, dipole, bending and pure shear, by imposing displacement of specific nodes,  as described in Fig.~\ref{fig:op_simulations}. We extract the coordinates (position, angle) of all the squares and use these to compute the dilation and shear field (method described in Appendix 1) and to calculate the relative amount of shear, as well as the accuracy of conformal map fits and of the analytic bulk-boundary correspondence (Fig.~\ref{fig:boundary_method}). Simulations are performed for a series of strains at identical material geometry as (i) and separately at linearly small strain across a range of hinge thicknesses $\ell = \{ 0.02, 0.033, 0.055, 0.092, 0.155, 0.258, 0.431, 0.719, 1.199, 2.0 \}$ mm.


\subsection*{Experimental Protocol \label{subsec:experiment}}

We 3D printed a elongated meta-beam of length 306mm, width 64mm and height 40mm, consisting of a checkerboard of lattice of $48\times 10$ squares of side $a=4.8$mm connected by hinges of thickness $\ell=0.2$mm using a Connex 500 stratasys 3D printed and the proprietary Stratatys Agilus 30 material (30 Shore A). This material is viscoelastic, its Young's modulus at short times is E= $3.3$ MPa and E= $0.6$ MPa at long times (see, e.g., Dykstra \textit{et al.} JAM 2019~\cite{Dykstra2019} for a calibration). For imaging purposes, we 3D printed black square-shaped pads ($3.6$mm) on one edge of the sample (proprietary Stratasys material: Vero Black). 

The sample was tested under two sets of non-uniform boundary conditions, (i) by a foot-shaped indenter, which was 3D printed in ABS Materials using a Dimension 3D printer (Stratasys); (ii) a three points bending (i.e. bridge) test, using a universal uniaxial mechanical testing machine (Instron 5943). In parallel to the mechanical loading, we recorded frames using a high-resolution camera ($6000\times4000$ pixels (Nikon D5600) with a telephoto lens (Nikon 200mm, F4), positioned at 4 meters from the sample. To ensure uniform lighting, the sample was illuminated by two led panels (Bresser, LG 900 54W) and white paper was used to ensure a uniform bright background. The 3D printed material tends to highly frictional and even adhesive, therefore, we used fine powder to lubricate the interaction with the bottom surface and the foot for (i). For (i), we performed a compression-decompression cycle at a rate of 0.1 mm/s up to a maximum 20 mm stroke, during which 400 frames were recorded (1 frame / sec). For (ii), we performed a compression-decompression cycle at a rate of 0.2 mm/s up to a maximum 40 mm stroke, during which 400 frames were recorded (1 frame / sec). The images were processed using standard image-tesselation and tracking techniques to extract the positions of the squares with subpixel detection accuracy of 0.02 pix (1 $\mu$m). We attribute the strongly different behavior observed between compression and decompression, to a combination of frictional, viscoelastic and self-adhesion effects.


\subsection*{Extracting Dilation and Shear Strains}
\label{sec:strain_measure}
To extract nonlinear measures of coarse dilation and shear strain magnitudes from deformation of RS-based structures, we track the square center displacements. Using the vectors connecting square centers to neighboring square centers as the infinitesimal material vectors in the reference $( \vvec{\dee x}^{(1)}, \vvec{\dee x}^{(2)})$ and target spaces $( \vvec{\dee X}^{(1)}, \vvec{\dee X}^{(2)})$, we may infer the deformation tensor $\ttens{F}_{ij} = \frac{\partial X_i}{\partial x_j}$. As the reference $\vvec{\dee x}^{(1)}$ and $\vvec{\dee x}^{(2)}$ are orthogonal, the recipe is simply
\begin{equation}\label{eq:def_tens_extract}
\ttens{F} = 
\begin{bmatrix}
    \frac{\vvec{\dee X}^{(1)}\cdot \vvec{\dee x}^{(1)}}{|\vvec{\dee x}^{(1)}|^2} & 
    \frac{\vvec{\dee X}^{(2)}\cdot \vvec{\dee x}^{(1)}}{|\vvec{\dee x}^{(2)}|^2} \\
    \frac{\vvec{\dee X}^{(1)}\cdot \vvec{\dee x}^{(2)}}{|\vvec{\dee x}^{(1)}|^2} & 
    \frac{\vvec{\dee X}^{(2)}\cdot \vvec{\dee x}^{(2)}}{|\vvec{\dee x}^{(2)}|^2}  
\end{bmatrix} \, .
\end{equation}
Then the induced metric of deformation is obtained via $\ttens{C} = \ttens{F}^T \cdot \ttens{F}$. Nonlinear dilation and shear strain magnitudes may be obtained from the invariants of the metric tensor via
\begin{equation}\label{eq:d_s_from_c}
  \begin{aligned}
    & d = \sqrt{\mathrm{Det}[\ttens{C}]} - 1, \\
    & s^2 =  \mathrm{Tr}[\ttens{C}] - 2 \sqrt{\mathrm{Det}[\ttens{C}]} \, .
  \end{aligned}
\end{equation}
These recipes, which are derived in greater detail in Appendix 1, give the dilation as the local area change, while the shear comes from the nonconformal part of the deformation as a complex map $h: z\rightarrow h(z, \bar{z})$, satisfying $s^2 = 4|\partial_{\bar{z}}h|^2$.

\subsection*{Fitting displacement data to the closest conformal map}
\label{sec:lse_method}

To identify the the conformal map $f(z)$ which most closely maps a set of $n_p$ points $\{z_i\}$ (expressed in complex form) to final positions $\{z'_i\}$, we minimize the error
\begin{equation}\label{eq:square_error}
Err \equiv \sum_{i=1}^{n_p} \left|z'_i - f(z_i) \right|^2  \, .
\end{equation}
Utilizing the analytic expansion of $f$ (equation~(\ref{eq:conformalmap})), we may find the set of coefficients $\{C_n\}$ which minimize the error by setting the partial derivatives of the error to zero, yielding equations
\begin{equation}\label{eq:solve_ls}
\sum_{n=1}^{n_c} A_{mn} C_n = \sum_{i=1}^{n_p}z'_i  \bar{z}_i^m  \, ,
\end{equation}

where we have defined the matrix
\begin{equation}\label{eq:define_A}
A_{mn} = \sum_{i=1}^{n_p} \bar{z}_i^m z_i^n \, 
\end{equation}
and the coefficients are now cut off at a maximum $n_c$. This effectively reduces the least-squares error analysis to a straightforward linear algebra problem, which is readily solved. We note that the cutoff $n_c$ should be chosen to be much less than the $n_p$ data points, to avoid over-fitting.  

To evaluate the accuracy of the fit, we define a similar function to the error in equation~\ref{eq:square_error}
\begin{equation}\label{eq:delta_func}
\Delta^2[u(z)] \equiv  \frac{  \sum_{i=1}^{n_p} |u_i - u(z_i)|^2  }{  \sum_{i=1}^{n_p} |u_i|^2  }   \, ,
\end{equation}
where $u_i = f_i - z_i$ is the observed displacement data, $u(z) \equiv f(z) - z$ are the displacements proposed to fit the data. The functional $\Delta^2[\,]$, quantifies the ``fraction of unexplained variance'' in the displacements, and constitutes a general method to quantify the amount of the deformation captured by a candidate conformal function, used in Fig.~\ref{fig:boundary_method}b,e as well as in the Appendices.


\subsection*{Inferring the interior conformal deformation from boundary dilation data}
\label{sec:boundary_method}

In our new bulk-boundary method for predicting deformation, we are able to infer the conformal deformation $f: z \rightarrow f(z)$ which will arise from a discrete set of $M$ applied dilations $\{ \alpha_k\}$ at a corresponding set of locations $\{z_k\}$ along the boundary of a simply-connected planar domain. Here, $\alpha_k = A'_k / A_k$ is ratio of local area elements before $A_k$ and after $A'_i$ deformation. The inference takes place in two steps: first an intermediate function $g(z) = log(f'(z))$ is obtained, then this function is used to integrate for the displacements. 

Because $f(z)$ will be conformal, so is the function $g(z) = log(f'(z))$ and therefore admits a similar expansion
\begin{equation} \label{eq:g_expand}
g(z) = \sum_{n=0}^{N-1} D_n z^n \, .
\end{equation}
where the cutoff $N$ should be significantly less than half the number of boundary points $M$ in order to avoid overfitting. Note now that since $g(z) = log(\alpha) + i \phi$, the real part of $g$ alone determines the dilation field. We may therefore infer the $g$ that fits the boundary conditions by minimizing the error
\begin{equation}\label{eq:g_err}
    \textrm{err} = \sum_k^M (\mathrm{Re}[g(z_k)] - \ln(\alpha_k) )^2 \, .
\end{equation}
Inserting equation~(\ref{eq:g_expand}), we minimize this error with respect to the coefficients $D_n = A_n + i B_n$ yielding the equations
\begin{equation}\label{eq:min_g}
    \begin{aligned}
    0 = & \frac{\partial [\textrm{err}]}{\partial A_l} = f^{(A)}_l + \sum_n^N A_n Q_{ln} + \sum_n^N B_n R_{ln} \\
    0 = & \frac{\partial [\textrm{err}]}{\partial B_l} = f^{(B)}_l + \sum_n^N A_n S_{ln} + \sum_n^N B_n T_{ln} \, ,
    \end{aligned}
\end{equation}
where
\begin{equation}\label{eq:min_g_mats}
    \begin{aligned}
    Q_{ln} = & \sum_k^M \left( r_k^{n+l} \mathrm{cos}(n\theta_k) \mathrm{cos}(l\theta_k) \right) \\
    R_{ln} = & -\sum_k^M \left( r_k^{n+l} \mathrm{sin}(n\theta_k) \mathrm{cos}(l\theta_k) \right) \\
    S_{ln} = & -\sum_k^M \left( r_k^{n+l} \mathrm{cos}(n\theta_k) \mathrm{sin}(l\theta_k) \right) \\
    T_{ln} = & \sum_k^M \left( r_k^{n+l} \mathrm{sin}(n\theta_k) \mathrm{sin}(l\theta_k) \right) \\
    f^{(A)}_{l} = & -\sum_k^M \left( \mathrm{ln}(\alpha_k) r_k^{l} \mathrm{cos}(l\theta_k) \right) \\
    f^{(B)}_{l} = & \sum_k^M \left( \mathrm{ln}(\alpha_k) r_k^{l} \mathrm{sin}(l\theta_k) \right)  \, , \\
    \end{aligned}
\end{equation}
and we have expressed $z_k = r_k e^{i\theta_k}$ in a complex polar form. equation~(\ref{eq:min_g}) reduces the inference of $g$ to a linear algebra problem which may be readily solved using built-in tools in, e.g., Mathematica. Noted also in Appendix 9, the row and column corresponding to $B_0 = \phi_0$ (the undetermined global rotation) are zero throughout and this row-column pair must be  removed before numerically solving. 

As described in Appendix 9, the function $g(z)$, specified by the coefficients $\{ D_n \}$, is used to reconstruct the $z$-derivative of the conformal map via $\partial_z f = \mathrm{Exp}(g)$. Generating the predictions of block displacements $u(z) = f(z) - z$, shown in main text Fig.~\ref{fig:boundary_method}d and evaluated in Fig.~\ref{fig:boundary_method}e,  is a matter of complex integration, which we accomplish using the built-in numerical integrators in Mathematica.

\section*{Acknowledgements}
We thank E. Matsumoto and M. Dimitriyev for insightful discussions and suggestions, S. Koot and D. Giesen for technical assistance. C.C. acknowledges funding from the
European Research Council via the Grant ERC-StG-Coulais-852587-Extr3Me. 

\section*{Data Availability} 

The experimental and finite element simulation data generated in this study have been deposited in the Zenodo database under accession code \url{https://doi.org/10.5281/zenodo.4646672}.

\section*{Code Availability} 
All the codes supporting this study have been deposited in the Zenodo database under accession code  \url{https://doi.org/10.5281/zenodo.4646672}.

\section*{Author Contributions}
M.D.C., C.C., M.v.H. and D.Z.R. together designed research, and contributed to writing. C.C. performed experiments and FEM simulations and processed data. M.D.C. performed boundary simulations, and analyzed simulation and experimental data. M.D.C. and D.Z.R. developed analytic theory. 

\section*{Competing Interests Statement}
The authors declare no competing interests.

\pagebreak

\appendix

\renewcommand{\thesubsection}{Appendix \arabic{subsection}}

\section*{Appendices}

\subsection{Quantifying Local Dilations and Shears \label{app:op_derive}}


To our knowledge, a parameter which quantifies what fraction of a \emph{nonlinear} deformation is composed of local dilation, as opposed to local shear, has not been introduced previously. We therefore present a more detailed derivation in the context of Finite Strain Theory. This is an essential ingredient for supporting our hypothesis in the main text that nonuniform deformations of a dilational metamaterial are nonetheless composed primarily of local dilations. 

The method described here applies to data describing deformation from a reference space $\vvec{X}$ to a target space $\vvec{x}(\vvec{X})$, which are smooth enough to take derivatives. In the case of the RS metamaterial, the square centers provide a discrete approximation to a smooth deformation, which may then be used to define local material derivatives. More generally, one may use the lattice vectors of a material with repeating metastructure as material vectors to take these material derivatives.  
From these derivatives we assemble the deformation tensor
\begin{equation}\label{eq:define_defTens}
F_{ij}(\vvec{x}) = \frac{\partial x_i}{\partial X_j} |_{\vvec{x}} \, .
\end{equation}
For a smooth deformation, this tensor contains all required information to quantify dilation and shear at each point in space.

Using the polar decomposition theorem, we may write $\ttens{F} = \ttens{R} \cdot \ttens{U}$, where $\ttens{R}$ is a rotation and $\ttens{U}$ is the symmetric "right stretch tensor". From this form, it is clear that an infinitesimal material vector in the reference space $\mathrm{d}\vvec{X}$ will deform first via multiplication with $\ttens{U}$, then will be rotated via $\ttens{R}$ to reach a final state $\mathrm{d}\vvec{x} = \ttens{R}\cdot\ttens{U} \cdot \mathrm{d}\vvec{X}$. Since $\ttens{U}$ is symmetric, there exists a local choice of coordinate system where $\ttens{U}$ is diagonal and acts on the local elements $\mathrm{d}\vvec{X}$ as a combination of just dilation and pure shear. The eigenvalues $(\lambda_1, \lambda_2)$ of $\ttens{U}$ are called the \textit{principal stretches} and the local material dilation and shear magnitudes may be written intuitively in terms of them. An infinitesimal square sample of side lengths $(\epsilon, \epsilon)$ oriented in the diagonal frame of $\ttens{U}$ will transform into a rectangle with side lengths $(\lambda_1 \epsilon, \lambda_2 \epsilon)$ after deformation. Using this, the local dilation may be defined intuitively in terms of the area
\begin{equation}\label{eq:dilation_def}
    d \equiv \frac{\Delta A}{A_0} = \mathrm{Det}[\ttens{U}] - 1 \, ,
\end{equation}
where $\Delta A$ is the change in area of our infinitesimal element and $A_0 = \epsilon^2$ is the initial area in the reference space. For a similarly intuitive notion of shear, we turn to the anisotropy of the tensor $\ttens{U}$.  We define the nonlinear shear magnitude in terms of the difference of the principal stretches as $s^2 = (\lambda_1 - \lambda_2)^2$. We will see that this definition matches with conventional intuition for standard simple shear nonlinearly as well as general mixed linear shear. Importantly, this shear may be rewritten in terms of invariants of $\ttens{U}$ as $s^2 = \mathrm{Tr}[\ttens{U}]^2 - 4 \mathrm{Det}[\ttens{U}]$, and is therefore also a rotationally invariant quantity.

\begin{figure*}[!t]  
\begin{center}
    \includegraphics[width=0.9\textwidth]{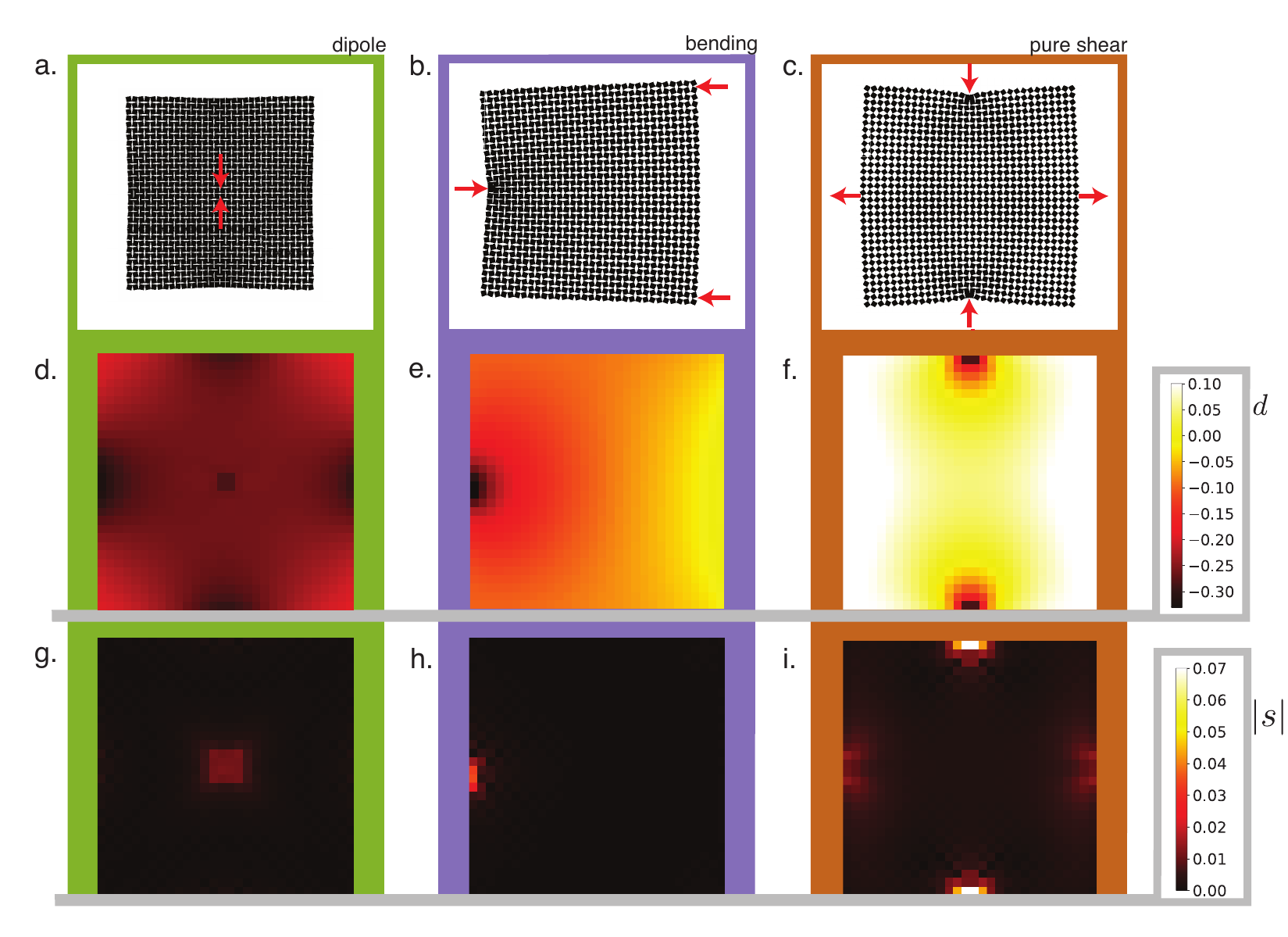}
    \caption{\label{fig:app_dilations_shears} 
Elastic RS-based structures deform with local dilation over shear.
For each of the applied displacement loads (in columns according to \textbf{a}, \textbf{b}, \textbf{c} ) from the main text, we display the spatial distributions of coarse dilation ($d$, \textbf{d}, \textbf{e}, \textbf{f}) and coarse shear magnitude ($|s|$, \textbf{g}, \textbf{h}, \textbf{i}) displaying a strong tendency of the RS-based metamaterial toward deformations that are locally dilation dominated. 
 }
    \end{center}
\end{figure*}

Since the stretch tensor is generally less convenient to extract numerically, we show that this local dilation and shear may also be written in terms of the trace and determinant of the metric (equivalently the ``right Cauchy-Green deformation tensor''). The metric is defined as
\begin{equation}
    C_{ij} = \frac{\partial x_k}{\partial X_i} \frac{\partial x_k}{\partial X_j} = (\ttens{F}^{\mathrm{T}}\ttens{F})_{ij}  = (\ttens{U}^{\mathrm{T}}\ttens{U})_{ij} \, .
\end{equation}
The coordinate system which diagonalizes $\ttens{U}$ also diagonalizes $\ttens{C}$, and it follows that the eigenvalues of the metric are the squared eigenvalues of $\ttens{U}$. Using this locally diagonalized frame, we may extract general relations between rotationally invariant quantities
\begin{equation}\label{eq:u_c_relate}
  \begin{aligned}
    &\mathrm{Det}[\ttens{U}]^2 =\mathrm{Det}[\ttens{C}], \\
    &\mathrm{Tr}[\ttens{U}]^2 = \mathrm{Tr}[\ttens{C}] + 2 \sqrt{\mathrm{Det}[\ttens{C}]}  \, .
  \end{aligned}
\end{equation}
Using these relations, we may write the more useful formulas
\begin{equation}\label{eq:app_d_s_from_c}
  \begin{aligned}
    & d = \sqrt{\mathrm{Det}[\ttens{C}]} - 1, \\
    & s^2 =  \mathrm{Tr}[\ttens{C}] - 2 \sqrt{\mathrm{Det}[\ttens{C}]} \, ,
  \end{aligned}
\end{equation}
completing our recipe to extract the local dilation and shear magnitudes from the metric tensor.


That these relations in equation~(\ref{eq:app_d_s_from_c}) discriminate conformal from non-conformal deformation is not entirely obvious. We therefore present them in alternative form by connecting to the complex formulation where the vector mapping $\vvec{x} \rightarrow \vvec{x}(\vvec{X})$ is captured by the complex function $z \rightarrow f(z, \bar{z})$. In this case, the components of the metric may be conveniently rewritten as
\begin{equation}\label{eq:right_cg_complex}
\begin{matrix}
C_{xx} = |\partial_z f|^2 + |\partial_{\bar{z}} f|^2 + 2 \mathrm{Re}\left[ \overline{\partial_z f} \partial_{\bar{z}} f \right] \\
C_{yy} = |\partial_z f|^2 + |\partial_{\bar{z}} f|^2 - 2 \mathrm{Re}\left[ \overline{\partial_z f} \partial_{\bar{z}} f \right] \\
C_{xy} = C_{yx} = 2 \mathrm{Im}\left[ \overline{\partial_z f} \partial_{\bar{z}} f \right]  \, ,
\end{matrix}
\end{equation}
where $\partial_z = 1/2(\partial_x - i \partial_y)$ and $\partial_{\bar{z}} = 1/2(\partial_x + i \partial_y)$ are standard complex derivatives~\cite{england2003complex}. With these relations, our metrics of conformal and non-conformal become
\begin{equation}\label{eq:extract_conf_metric2}
\begin{aligned}
d^2 &= \left[|\partial_z f|^2 - |\partial_{\bar{z}} f|^2 - 1 \right]^2,  
\\ \nonumber
s^2 &= 4 |\partial_{\bar{z}} f|^2 \, ,
\end{aligned}
\end{equation}
where, remembering that the conformal condition is $\partial_{\bar{z}}f=0$, the shear part now more transparently quantifies the non-conformal part of the deformation. While both dilation and shear are strain quantities, there is some ambiguity in comparing them nonlinearly. The expressions used here, composed entirely in terms of invariants of $\mathbf{C}$, are constructed to be intuitive and geometric.
In addition, rewriting
\begin{equation}\label{eq:shears2}
    s^2 =  (\partial_x u_x - \partial_y u_y)^2 + (\partial_x u_y + \partial_y u_x)^2 
\end{equation}
in terms of the components of the displacement $\vvec{u} = \vvec{x} - \vvec{X}$ reveals that our definition of the shear matches with the combined magnitude of pure and simple shears (first and second terms in equation~\ref{eq:shears2} respectively) from linear strain theory. Note that, in the linear limit, these strain measures satisfy $1/2 (d^2 + s^2) = \epsilon_{ij}\epsilon_{ij}$. While alternate definitions of shear may be identified which also satisfy this in the linear limit, we have explored these and they do not qualitatively change the results presented here. Finally, for a conventional simple shear $\vvec{u} = \gamma y \hat{x}$, our formula reassuringly recovers $s^2 = \gamma^2$ to nonlinear order.

As shown in Fig.~\ref{fig:app_dilations_shears}, these formulae reveal that the typical magnitude of local dilation is much larger than that of local shear. While the shear may become large in small regions near the applied loads and where the dilation changes over short distances, dilation still dominates by nearly an order of magnitude at the worst. Nonuniform, nonlinear deformations of the RS-based mechanism in force-balance may then indeed be composed of local dilations with comparatively very little shear.





\subsection{Least-squares method to extract the nearest conformal map from data\label{app:lse_extract}}

We now identify a simple fitting method of locating the closest conformal map describing a discrete set of displacement data points. Per \ref{app:op_derive}, the discrete data should be chosen carefully to approximate a smooth deformation. For lattice materials, this means tracking identical points in neighboring unit cells. For the RS-based metamaterial, we can choose the centers of the square elements.   

The method consists of first noting that a general conformal map $f(z)$ on a simple finite planar domain with no holes (i.e. simply-connected) may be expressed precisely as an analytic function with a series expansion
\begin{equation}\label{eq:conformal_expand_app}
f(z) = \sum_{n=0}^\infty C_n z^n \, .
\end{equation}
In contrast, a generic non-conformal deformation must be expressed as a double sum over all products of $z$ and the complex conjugate $\bar{z}$. 
The complex expansion coefficients $C_n$ provide a complete description of the map and solving for them is therefore the goal of our analysis. Finding the nearest conformal map is a matter of minimizing the square error function
\begin{equation}\label{eq:app_square_error}
Err \equiv \sum_{i=1}^{n_p} \left|f_i - f(z_i) \right|^2  \, ,
\end{equation}
where the sum runs over the $n_p$ discrete data points $z_i \rightarrow f_i$. Inserting the form of $f(z)$ (equation~(\ref{eq:conformal_expand_app})) and minimizing with respect to the $C_n$, we find
\begin{equation}\label{eq:app_solve_ls}
\sum_{n=1}^{n_c} A_{mn} C_n = \sum_{i=1}^{n_p}f_i  \bar{z}_i^m  \, ,
\end{equation}
%
%
where we have defined the matrix
\begin{equation}\label{eq:app_define_A}
A_{mn} = \sum_{i=1}^{n_p} \bar{z}_i^m z_i^n \, 
\end{equation}
and the coefficients are now cut off at a maximum $n_c$. This effectively reduces the least-squares error analysis to a straightforward linear algebra problem, which is readily solved. We note that the cutoff $n_c$ should be chosen to be much less than the $n_p$ data points, to avoid over-fitting.  

\begin{figure}[!t]  
\begin{center}
    \includegraphics[width=0.45\textwidth]{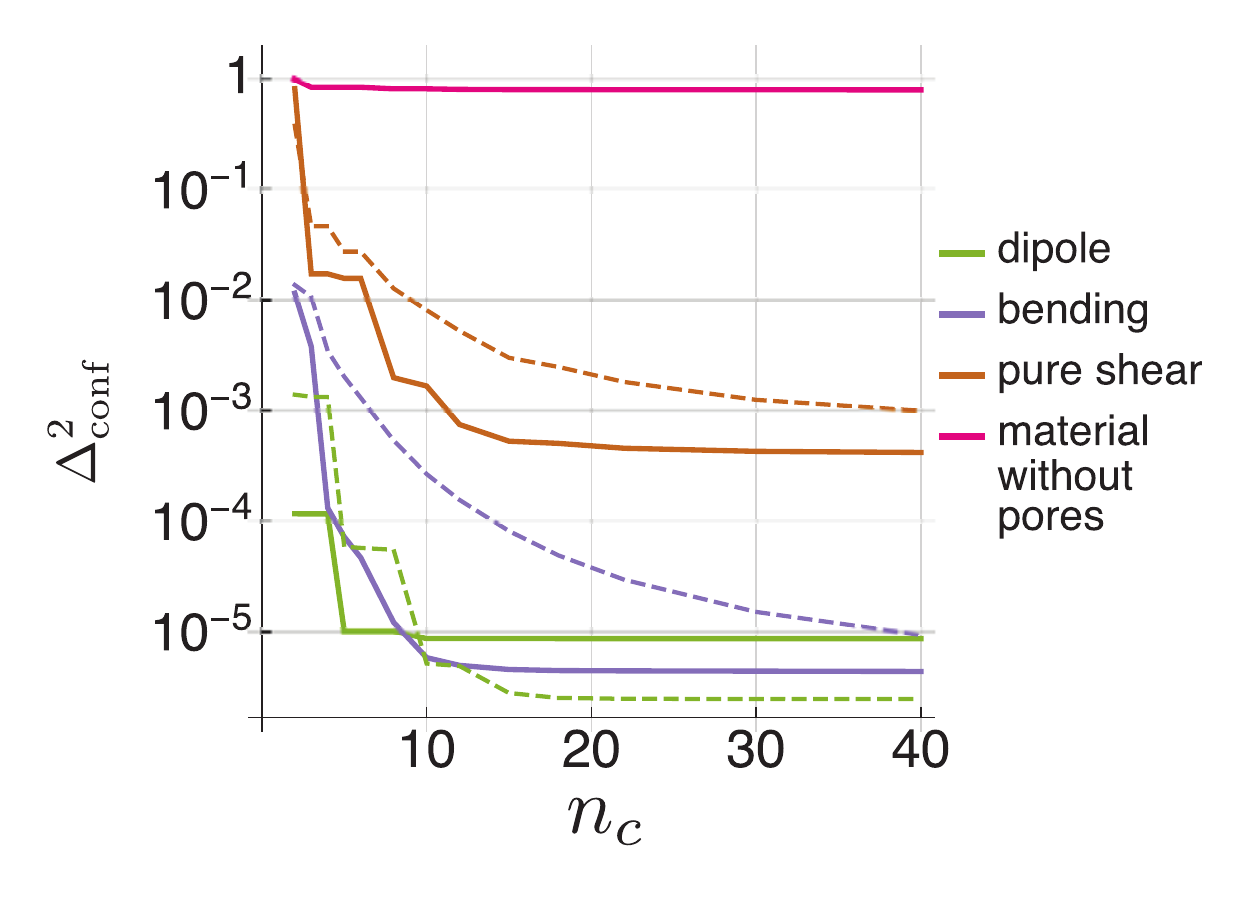}
    \caption{\label{fig:app_lse_coefficients} The accuracy of the conformal map with the increasing number of coefficients allowed in the expansion equation~(\ref{eq:conformal_expand_app}). Solid lines correspond to fits with FEM data at the smallest strains, while the dotted lines are for the largest strains explored in the main text analysis. In comparison to the conventional elastic material in the absence of pores, the fitting quickly reaches very small errors $\Delta_{\text{conf}}^2$ which is the function $\Delta^2[\,]$ evaluated for the particular candidate fit $u(z)$. For loads with discrete rotational symmetry, the errors go down in steps, since only every few coefficients may be non-zero.}
    \end{center}
\end{figure}

To evaluate the accuracy of the fit, we define a function
\begin{equation}\label{eq:app_delta_func}
\Delta^2[u(z)] \equiv  \frac{  \sum_{i=1}^{n_p} |u_i - u(z_i)|^2  }{  \sum_{i=1}^{n_p} |u_i|^2  }   \, ,
\end{equation}
where $u_i = f_i - z_i$ is the observed displacement data, $u(z) \equiv f(z) - z$ are the displacements proposed to fit the data. The functional $\Delta^2[\,]$, known as the fraction of variance unexplained, is then a general method to quantify the amount of the deformation captured by a candidate conformal function, which we use in the main text and in other appendices. The method evaluates error using the displacements rather than positions in order to avoid a false inflation of the results at small deformation. Inserting the numerically solved closest fit into equation~(\ref{eq:app_delta_func}), we show in Fig.~\ref{fig:app_lse_coefficients} that only around 20 coefficients are needed to fit the deformations of the RS metamaterial before the improvement starts to become negligible. This is good, as the more coefficients that are included, the more computationally expensive the fitting problem becomes.

We remark that, in practice, application of this method works best if the coordinate system is first centered and rescaled so that the data points reside closer to unity away from the origin. This helps to avoid large and small number issues in the numerics, which includes higher and higher exponents of the positions as more coefficients are included.

This constitutes a fast and accurate method to extract the nearest conformal map from position data. Note that, while the method is reduced to linear algebra, no assumptions of linearity of the strains in either the deformation, or the fitting map, have been made. Indeed, this method shows the nonlinear deformations of the RS metamaterial are well-approximated as a conformal map to nonlinear order.


\subsection{Energetics of Hinge Deformations}\label{app:hinges}


We now aim to derive a continuum elasticity theory for the RS metamaterial. However, to consider the RS elastic material at full detail requires allowance of all possible deformations of the porous elastic structure. Such a tensor field theory for deformations would be complicated and unwieldy, leaving very little room for analytic progress. Fortunately, due to the careful design of the pores, we may think of the material as composed of square elements of side length $a$ connected by comparatively small ``hinge'' elements of thickness $\ell$. This enables an effective description of the material as a collection of hinges, written in terms of simplified variables which prove to be very convenient for coarse-graining. We establish this effective ``hinge-based'' energetic formalism below.

\begin{figure*}[!t]  
\begin{center}
    \includegraphics[width=0.85\textwidth]{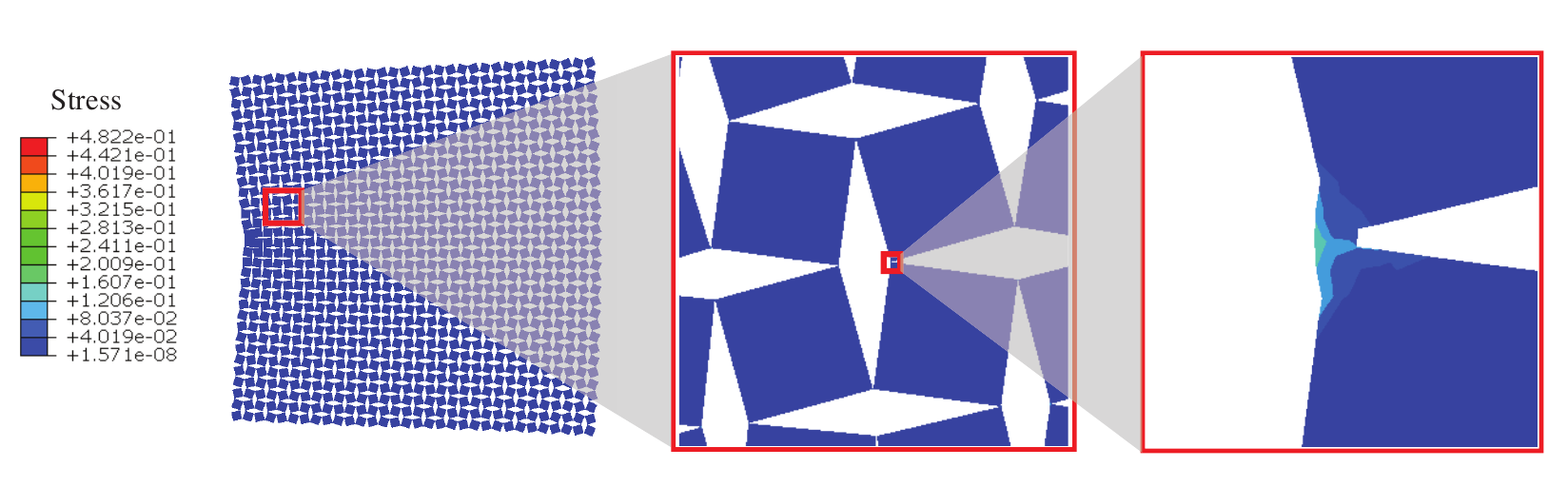}
    \caption{\label{fig:hinge_localization}  Sample deformations from Finite Element Simulations displaying the detail of the elastic mesh method as well the characteristic localization of the stress (and therefore also strain) magnitude to a small region in the vicinity of the hinge. }
    \end{center}
\end{figure*}

In the limit that $\ell \ll a$, the hinge will become very flexible compared to the stiff square pieces. We therefore follow successful previous examples~\cite{Coulais2018, Deng2019} in which the large square elements are approximated as rigid bodies and all strain deformation is assumed to take place at the hinges. This assumption is readily verified in the Finite Element deformation data, where the stress is generically found to be dramatically localized to the hinge region, as shown in  Fig.~\ref{fig:hinge_localization}.  

\begin{figure*}[!t]  
\begin{center}
    \includegraphics[width=0.85\textwidth]{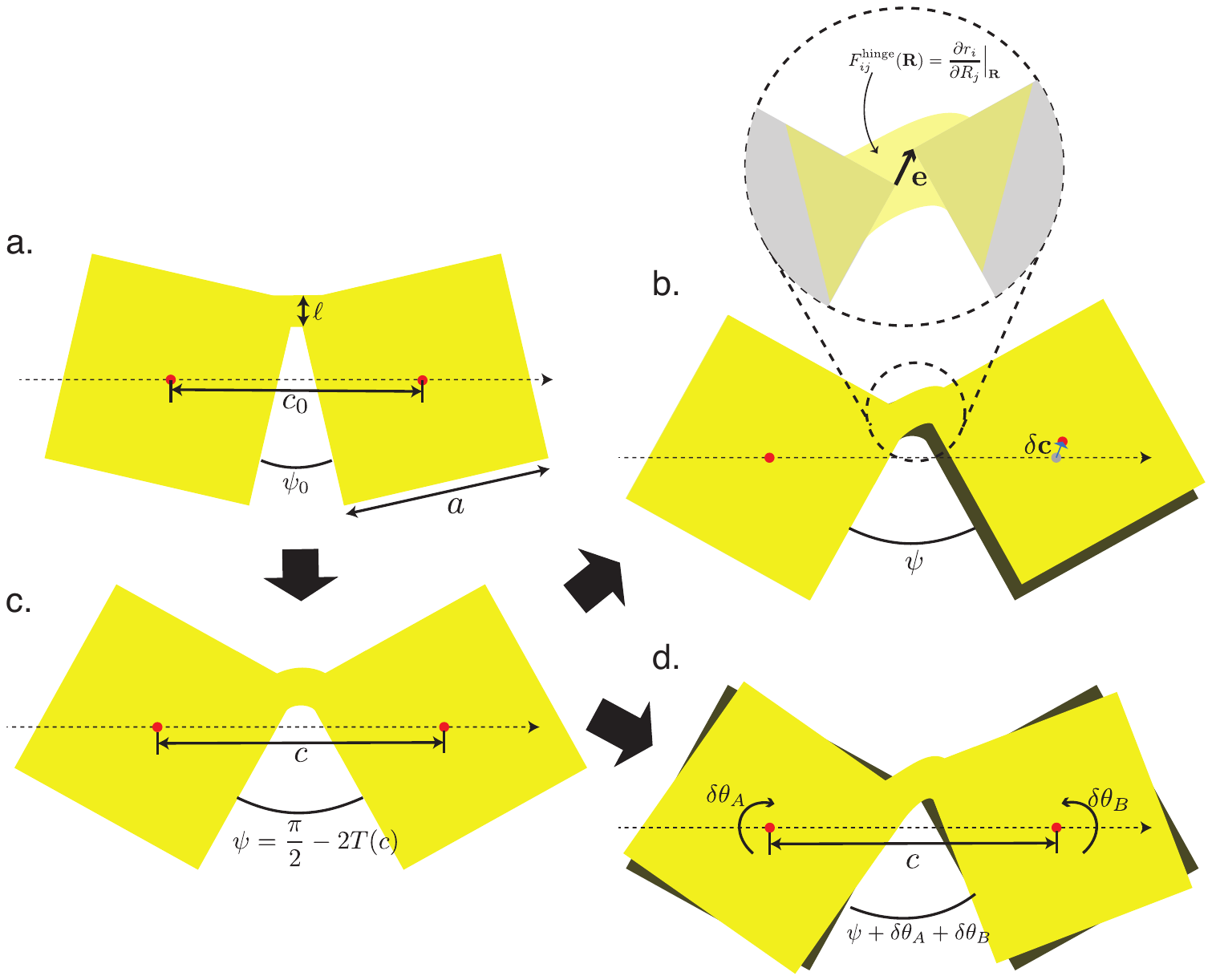}
    \caption{\label{fig:hinge_deformations} In the absence of additional forces applied near the hinge, the elastic hinge energetics may be parametrized with three numbers describing the relative orientations and positions of the squares. First, the ``mechanism'' portion (\textbf{a} $\rightarrow$ \textbf{c}) may be traversed by fixing the center-to-center distance $c_0 \rightarrow c$ of the squares and allowing the square orientations to relax. Then small deviations around this state are allowed by either (\textbf{d}) preserving the square distance and allowing small orientation changes ($\delta \theta_A, \delta \theta_B$) or (\textbf{b}) preserving the square orientations and allowing a small change ($\delta\vvec{c}$) in relative square positions. }
    \end{center}
\end{figure*}

To quantify the energetics in this effective description, we consider a single hinge connecting two elastic blocks as shown in Fig.~\ref{fig:hinge_deformations}-a. Knowledge of the deformation tensor $F^{\text{hinge}}_{ij}(\vvec{r})$ throughout the hinge completely determines the system configuration, as stress and strain are already assumed to be zero elsewhere. Here, $\vvec{r}$ and $\vvec{R}$, respectively, are target and reference coordinate systems describing the fine structure of the hinge deformations, in contrast to $\vvec{x}$ and $\vvec{X}$ which will describe a coarsened picture of continuum deformations of the metamaterial at the mesoscale. Given the complete constitutive relation for the elastic material, the tensor field $F^{\text{hinge}}_{ij}(\vvec{r})$ provides sufficient information to compute the energy density at each point in space. For generality we write this simply as an unknown function  $e(\vvec{r}) = \phi(F^{\text{hinge}}_{ij}(\vvec{r})) $. 



We now confine our focus to hinges with aspect ratio $\sim 1$, which is the case for the experiments and simulations in this manuscript. This condition generally removes opportunities for bistability and other path-dependent elastic behavior at the hinge. Therefore, prescribing the orientations and  positions of the two squares attached to a particular hinge will leave only one possible force-balanced deformation $F^{\text{hinge}}_{ij}(\vvec{r})$ in the hinge. The energy of this hinge may therefore be rewritten as a function of the two square orientations and two 2d vector positions, six numbers altogether. Accounting for translational and rotational symmetry leaves only three scalar parameters necessary to describe the configuration of the hinge. We will focus on two distinct, yet convenient choices for these three parameters as described below. 

Importantly, we note that arbitrary relative positions and orientations of a pair of squares will generally cost a great deal of energy and may stretch the hinge to the point of fracture. We wish to focus on low-energy configurations that do not stretch the hinge too much, which includes a particular set of nonlinear rearrangements due to the design based on an ideal-hinge mechanism. Therefore, in order to find an appropriate parametrization for these low-energy configurations, we must first quantify the finite-hinge (finite-energy) analogue of this mechanism.
As shown in Fig.~\ref{fig:hinge_deformations}a,b, fixing the distance between the square centers $(\cc)$ and allowing the square orientations to relax defines our twisting angle $T(\cc) \equiv \pi/4 - \psi(\cc)/2$. Matching with the common nomenclature for twisting in kagome and RS lattices, this represents the amount the squares must each be rotated away from the fully extended state to balance torque at fixed distance.  
To twist the squares oppositely according to $T(\cc)$ while moving the square centers to distance $\cc$ therefore constitutes our generalized mechanism. In the limit that the hinge size $\ell$ goes toward zero, and the strain is increasingly localized to an infinitesimal hinge, this motion must approach the ideal frictionless hinge version of the mechanism. We may therefore write the twisting function as
\begin{equation}\label{eq:real_mechanism}
T(\cc) = \mathrm{ArcCos} \left( \frac{\cc}{\sqrt{2} a} \right) + \frac{\ell}{a} \delta_T(\cc) + \mathcal{O}(\ell/a)^2 \, 
\end{equation}
where $\delta_T(\cc)$ is an unknown dimensionless function capturing the first-order corrections to the twisting due to the finite size and stiffness of the hinge. Eventually we will proceed to ignore this first order correction, but must acknowledge that the function $\delta_T(\cc)$ may diverge in the vicinity of $\cc=\sqrt{2}a$ (the untwisted state). Both our hinge theory and subsequent coarse-graining of the Rotating Square (RS) lattice mechanics will break down near this untwisted state and we therefore leave analysis of this anomalous point in the configuration space for future work. 

Alternatively, we may explore this mechanism by fixing the orientation $\psi = \pi/2 - 2 T$ and allowing the positions to relax to a square center spacing $\cc(T)$ which can be expressed as
\begin{equation}\label{eq:real_mechanism2}
    \cc(T) = \sqrt{2} a \mathrm{Cos}(T) + \ell \delta_{\cc}(T) \, .
\end{equation}
Again, the second term  on the right-hand side is a correction term dependent on an unknown function $\delta_{\cc}$ which will eventually be neglected in the small hinge limit. 

We now use these mechanism twisting functions to define two parametrizations of the hinge energy. In both cases, we first choose to move to a point along the mechanism pathway, setting the distance $\cc$ and twisting angle to $\psi = \pi/2 - 2 T(\cc)$ as shown in Fig.~\ref{fig:hinge_deformations} (left). We may then deviate from this configuration in two ways: In the first, we may choose to fix the square orientations and allow a small vector deviation $\vvec{\delta \cc}$ in the relative positions as shown in the top right of Fig.~\ref{fig:hinge_deformations}. We will refer to this as the ``vector parametrization''. Alternatively, we could keep the square centers fixed in place and allow each to have a small reorientation away from the mechanistic twisting orientation as shown in Fig.~\ref{fig:hinge_deformations}. This defines deviation angles $\delta \theta_A$ and $\delta \theta_B$ and we will refer to this as the ``angular parametrization''. In the end, the choices will span the same configurations of the hinge structure. 

Both parametrizations above are useful for isolating the ``non-mechanistic'' part of the hinge motion into small variables. It is clear for small hinges that comparatively large $\vvec{\delta \cc}$ will put the hinge under large strains which scale roughly as $\sim \frac{|\delta \cc|}{\ell}$ and will put conventional materials at risk of tearing and fracture. We therefore work in the limit $\frac{|\delta \cc|}{\ell} \ll 1$. Equivalently, $(\delta \theta_A, \delta \theta_B) \ll 1$ are the appropriate conditions in the angular parametrization. We will see below that the vector parametrization is useful for deriving scaling relations, while the angular parametrization is useful for the coarse-graining. Given some configuration of the hinge we may now write the hinge energy in terms of the vector parameters
\begin{equation}\label{eq:vec_param_energy}
    E_H = g_v(T, \vvec{\delta \cc}, \ell, a, T_0) \, ,
\end{equation}
or equivalently in terms of the angular parameters
\begin{equation}\label{eq:ang_param_energy}
    E_H = g_a(\cc, \delta \theta_A, \delta \theta_B, \ell, a, T_0) \, .
\end{equation}
where $T_0$ is the ``pretwist'' present in the reference configuration of the hinge ensemble. 
Finally, we note that it is useful to be able to convert between these descriptions. Neglecting second and higher order terms in the deviations $\vvec{\delta \cc}$ and dimensionless hinge size $\ell/a$, we may write
\begin{equation}\label{eq:param_convert}
\begin{aligned}
\delta \theta_A = -\frac{(\delta \cc)_x}{\sqrt{2 a^2 - \cc^2}} + \frac{(\delta \cc)_y}{\cc}    \\
\delta \theta_B = -\frac{(\delta \cc)_x}{\sqrt{2 a^2 - \cc^2}} - \frac{(\delta \cc)_y}{\cc} \, .
\end{aligned}
\end{equation}

Even with perfect knowledge of the hinge geometry and nonlinear material elasticity, finding the form of the functions $g_v, g_a$ would require solving a difficult nonlinear mechanics problem. However, a lot can be discerned \emph{without} knowing the exact form of the energy and instead relying only on scaling arguments. In the vector parametrization, our hinge deformation tensor is a function not only of the location within the hinge $\vvec{x}$, but also this tensor field must depend on the inputs $T$ and  $\vvec{\delta \cc}$. It is instructive to consider rescaling the hinge reference and final states by a factor $\gamma$ while preserving shape. This means that points in the hinge $\vvec{r}$ are mapped to points $\tilde{\vvec{r}} =  \gamma \vvec{r}$ and the deformation tensor is preserved  via $\tilde{F}^{\text{hinge}}_{ij}(\tilde{\vvec{r}}) = F^{\text{hinge}}_{ij}(\vvec{r})$. Because the deformation tensor is preserved, so is the energy density. The only difference in computing the total energy of the rescaled hinge comes from integration over a rescaled domain, and we find $\tilde{E_H} = \gamma^2 E_H$. However, this rescaling also impacts the relative locations of the large square elements. Consider the internal vector $\vvec{e}$ which points from the projected corner of square $A$ to the projected corner of square $B$ as shown in the zoom in Fig.~\ref{fig:hinge_deformations}-b. Due to the preserved shape of the hinge, this internal vector must also scale directly with $\gamma$. This internal vector also has an analytic form
\begin{equation}\label{eq:hinge_extension}
\vvec{e} = \ell \delta_{\cc}(T) \hat{x} + \vvec{\delta \cc} \, ,
\end{equation}
where $\delta_{\cc}(T)$ is again the correction to the mechanism from equation~(\ref{eq:real_mechanism2}). Considering the configurations where $\vvec{\delta \cc}=0$ in equation~(\ref{eq:hinge_extension}), we note that because $\vvec{e}$ scales with $\gamma$, and because $\ell$ also scales directly with $\gamma$, then $\delta_T$ must not scale with $\gamma$. Noting that $\delta_T$ is independent of $\vvec{\delta d}$, it will remain scale-free when we also apply such a displacement $\vvec{\delta \cc}$. Because each side of equation~(\ref{eq:hinge_extension}) must scale the same way with $\gamma$, we have now established the result that the displacement vector $\vvec{\delta \cc}$ indeed must scale linearly with $\gamma$. 

Rescaling the hinge has no impact on the square relative orientations $\psi$ nor on the ``pre-twist'' $T_0$. This rescaling impacts only $\ell$ and $\vvec{\delta \cc}$ and the hinge total energy $E_H$ as described above (in the vector parametrization). We may therefore rewrite the hinge energy function
\begin{equation}\label{eq:vec_rescale_energy0}
    E_H = g_v(T, \vvec{\delta \cc}, \ell, a, T_0) = \ell^2 g_1(T,\frac{ \vvec{\delta \cc}}{\ell}, a, T_0)  \, ,
\end{equation}
where $g_1$ is another unknown function, whose value may be thought of as a mean energy density in the hinge. 

Another way of arriving at equation~(\ref{eq:vec_rescale_energy0}) is the following: given we had started with a smaller hinge of the same initial shape, we must arrive at the same final force-balanced hinge shape by simply rescaling the applied displacement vector $\vvec{\delta \cc}$ by the same factor as $\ell$. Therefore, we must be able to write the hinge energy as in equation~(\ref{eq:vec_rescale_energy0}) as a function of the current twist, the pretwist, and the value of $\vvec{\delta \cc}$ relative to $\ell$.

We now apply our previous assumption that we will avoid extreme, high energy configurations of the hinge where material fracture becomes likely. Quantitatively, this is the condition $\frac{\vvec{\delta \cc}}{\ell} \ll 1$. We may use this condition to expand the energy in equation~(\ref{eq:vec_rescale_energy0}) to second order, yielding
\begin{equation}\label{eq:vec_rescale_energy}
\begin{aligned}
    E_H = & \ell^2 g_T(T)  \\ & + \frac{(\vvec{\delta \cc} \cdot \hat{\cc})^2}{2} k_s(T) + \frac{(\vvec{\delta \cc} \cdot \hat{\cc}_{\bot})^2}{2} k_\mu(T) \\ & +  \mathcal{O}\left(\frac{\vvec{\delta \cc}}{\ell}\right)^3  \, ,
    \end{aligned}
\end{equation}
%
The twist-dependent coefficients may be written in terms of $g_1$ via
\begin{equation}\label{eq:e_coeffs_vec}
\begin{aligned}
g_T(T) & = g_1(T, 0, T_0) \\
k_s(T) & = \frac{\partial^2 g_1}{\partial (\vvec{\delta \cc} \cdot \hat{\cc})^2}\bigg\rvert_{T, \vvec{\delta \cc} = 0, T_0} \\
k_\mu(T) & = \frac{\partial^2 g_1}{\partial (\vvec{\delta \cc} \cdot \hat{\cc}_\bot)^2}\bigg\rvert_{T, \vvec{\delta \cc} = 0, T_0} \, .
\end{aligned}
\end{equation}

\noindent Using equations~\ref{eq:param_convert}~\&~\ref{eq:e_coeffs_vec}, we can convert this energy to the angular parametrization. This becomes
\begin{equation}\label{eq:ang_rescale_energy}
    \begin{aligned}
    E_H = & \ell^2 g(\tilde{\cc})  \\ & +  \frac{ a^2 k_1(\tilde{\cc}) }{2} ( \delta \theta_A^2 + \delta \theta_B^2) + a^2 k_2(\tilde{\cc}) \delta \theta_A \delta \theta_B \\ & + \mathcal{O}\left(\frac{\vvec{\delta \cc}}{\ell}\right)^3  \, ,
    \end{aligned}
\end{equation}
where $\tilde{\cc} = \frac{\cc}{\sqrt{2}a}$ is the dimensionless rescaling of the square center distance and the moduli are defined by
\begin{equation}\label{eq:e_coeffs_ang}
\begin{aligned}
g(\tilde{\cc}) = & \, g_T(T(\sqrt{2}a \tilde{\cc})) \\  & =  g_T(\mathrm{ArcCos}(\tilde{\cc} \mathrm{Cos}(T_0))) \\
k_1(\tilde{\cc}) = & \frac{\mathrm{Sin}(T(\sqrt{2}a \tilde{\cc}))^2}{2} k_s(T(\sqrt{2}a \tilde{\cc})) \\ & + \frac{\mathrm{Cos}(T(\sqrt{2}a \tilde{\cc}))^2}{2} k_\mu(T(\sqrt{2}a \tilde{\cc})) \\
k_2(\tilde{\cc}) = & \frac{\mathrm{Sin}(T(\sqrt{2}a \tilde{\cc}))^2}{2} k_s(T(\sqrt{2}a \tilde{\cc})) \\ & - \frac{\mathrm{Cos}(T(\sqrt{2}a \tilde{\cc}))^2}{2} k_\mu(T(\sqrt{2}a \tilde{\cc})) \, .
\end{aligned}
\end{equation}
While the function $g_1$ remains unknown, and the energies in equations~\ref{eq:ang_param_energy}~\&~\ref{eq:vec_param_energy} may not be evaluated directly, these expressions provide useful insight into the nonlinear macroscopic energetics of the RS metamaterial while excluding extremal configurations, as shown in \ref{app:coarse_grain}.
  Importantly, as the function $g_1$ defined in equation~(\ref{eq:vec_rescale_energy0}) is scale-free, there is no remaining ``hidden'' dependance on the hinge size $\ell$ nor the square size $a$ in the energy function and the scaling is directly apparent.
  
  We have therefore avoided a complicated tensor-field description of our structured elastic metamaterial, in favor of a hinge-based energetic formalism based on a greatly simplified set of degrees of freedom: the positions and orientations of the square elements. In this formalism, energetics and scaling permit analytic descriptions which prove helpful for coarse graining.


\subsection{Coarse-Graining Procedure for Nonlinear Nonuniform Mechanics of the Rotating Square Lattice}\label{app:coarse_grain}


We now derive  the continuum elasticity theory for the rotating-square (RS) lattice, as presented in the main text. Following \ref{app:hinges}, we approximate the elastic RS metamaterial as a collection of rigid square elements connected by comparatively  flexible hinges. The hinges and square elements may be made out of the same material, while the comparative flexibility arises due to the small size of the hinge. 

We establish the local continuum energy penalty created by a mapping $\vvec{x} \rightarrow \vvec{X}(\vvec{x})$. This mapping will control the location of the square centers, while the orentations are left free to realize torque balanced configurations. Following the main text result that soft deformations of the RS metamaterial are near-conformal, we restrict our focus to deformations composed of a large nonlinear dilation, a finite local rotation, and a small generic shear. This prescription is captured by the deformation tensor
\begin{equation}\label{eq:app_deftens}
    F_{ij}(\vvec{x}) = R(\phi(\vvec{x}))_{ik} \alpha(\vvec{x}) \left[ \delta_{kj} + \frac{s_1(\vvec{x})}{2} \ttens{\sigma}^{(3)}_{kj} + \frac{s_2(\vvec{x})}{2} \ttens{\sigma}^{(1)}_{kj} \right]
\end{equation}
where 
\begin{equation}\label{eq:pauli}
    \ttens{\sigma}^{(1)} = 
    \begin{bmatrix}
        0 & 1 \\
        1 & 0 
    \end{bmatrix}
    \qquad \& \qquad
    \ttens{\sigma}^{(3)} = 
    \begin{bmatrix}
        1 & 0 \\
        0 & -1 
    \end{bmatrix}
\end{equation}
are Pauli matrices and $\ttens{R}(\phi)$ a standard rotation by $\phi$. 
To understand this tensor, we break it down piece by piece. First, recall that by definition this tensor describes the Jacobian of the deformation map $\vvec{X} \rightarrow \vvec{x}(\vvec{X})$ via $F_{ij} = \frac{\partial x_i }{\partial X_j}$. The infinitesimal material vectors are therefore understood to transform under deformation  via $\dee\vvec{X} \rightarrow \dee\vvec{x} = \ttens{F} \cdot \dee\vvec{X}$. Consider, then, the action of each element of $\ttens{F}$ going from right to left, in the order that they act on $\dee\vvec{X}$. The first (rightmost, in square brackets) term applies a linear strain composed solely of shears $s_1(\vvec{X})$ and $s_2(\vvec{X})$ defined as
\begin{equation}\label{eq:def_s1s2}
\begin{aligned}
s_1 = \partial_x u_x - \partial_y u_y \\
s_2 = \partial_x u_y + \partial_y u_x
\end{aligned}
\end{equation}
where $u_i$ is the $i$-th component of displacement, defined at this intermediate deformation for the purposes of understanding. Defined in this way, $s_1$ and $s_2$ correspond to the traditional notions of pure shear and simple shear, respectively. Following the application of the linear shear, the material vectors undergo an isotropic dilation $\alpha(\vvec{X})$ and are then rotated by $\phi(\vvec{X})$, bringing us to the final configuration. While this is a general form for such a (large dilation plus small shears) deformation, we note that these new variables $\phi, \alpha, s_1, s_2$ cannot vary arbitrarily through space, as they must still obey a closure condition to guarantee geometric (mechanical) compatibility. This is because a generic, spatially varying candidate strain field may not correspond to a valid displacement field, just as a generic candidate electrical field does not always correspond to an electrostatic potential. In the vanishing shear limit the closure condition takes the form
\begin{equation}\label{eq:app_conf_closure}
0 = \partial_i \alpha - \alpha \epsilon_{ij} \partial_j \phi \, ,
\end{equation}
 where $\epsilon_{ij}$ is the 2-by-2 antisymmetric unit tensor. Reassuringly, equation~(\ref{eq:app_conf_closure}) captures the Cauchy-Riemann condition for a deformation to be conformal. To see this, note that the derivative of a conformal function may be written in modulus-argument form as $\partial_z f = f' = \alpha \exp[i\phi]$. Then, as conformal functions must be functions of $z$ only, we must have $\partial_{\bar{z}} f' = 0$. Writing out the real and imaginary parts of this condition leads to the x- and y-components of equation~(\ref{eq:app_conf_closure}). 
 

We now choose the vectors $\vvec{\cc}_\nu$ connecting neighboring square centers (and where $\nu = \{1, 2, 3, 4\}$ indexes \emph{which} neighbor) to represent the infinitesimal vector elements of the effective continuum material, and thus $F_{ij}$ prescribes all the relative square locations after deformation. If we can now find a similar recipe for writing down the square orientations in terms of continuum quantities, then our description of the metamaterial configuration is effectively complete. We will then be able to write down the hinge energies and sum these over the unit cell to get a continuum energy density. To identify the torque-balanced square orientations, we note that there are only two squares per unit cell. We therefore assume that square orientations are controlled by two angular fields which vary slowly through space.  A unit cell contains squares $A$ and $B$ which after deformation are located at $\vvec{r}_A$ and $\vvec{r}_B$ and have orientations $\theta_A$ and $\theta_B$. There is already some intuition for the expected orientations in terms of the counter-rotation mechanism of the squares arising from the local lattice dilation factor $\alpha(\vvec{X})$. This is expressed by writing the orientations as 
\begin{equation}\label{eq:orientation_fields}
    \begin{aligned}
    \theta_A = T(\sqrt{2} a \alpha(\vvec{r}_A)) + \phi(\vvec{r}_A) + \Delta T(\vvec{r}_A) + \Delta \phi(\vvec{r}_A) \\
    \theta_B = - T(\sqrt{2} a \alpha(\vvec{r}_B)) + \phi(\vvec{r}_B) - \Delta T(\vvec{r}_B) + \Delta \phi(\vvec{r}_B) \, ,
    \end{aligned}
\end{equation}
where $T()$ is the same mechanism function from \ref{app:hinges}, while $\Delta T, \Delta \phi$ are the two smooth fields required to capture the two orientation degrees of freedom in each unit cell. Note that while equation~(\ref{eq:orientation_fields}) relies on a well-motivated guess, there is no loss of generality due to the presence of the correction fields $\Delta T, \Delta \phi$, and conveniently that if our guess is correct, then enforcing torque balance will simply set the value of these fields locally to zero. In the limit of vanishing shears and vanishing gradients of $\alpha$ and $\phi$, we find that $\Delta T$ and $\Delta \phi$ also vanish and therefore the magnitude of these angular correction fields must be on the order of the shears or $\nabla \phi$ or smaller.

We are now equipped with a sufficient description to specify the energy (equation~\ref{eq:ang_rescale_energy}) of an arbitrary hinge connecting the square initially centered at $\vvec{X}$ to the square initially centered at $\vvec{X} + \vvec{\cc}^0_\nu$ in terms of our continuum quantities $(\phi, \alpha, s_1, s_2, \Delta T, \Delta \phi)$. 
Constructing the required geometric quantities to evaluate the hinge energy, we first find via application of the deformation tensor that the length of $\vvec{\cc}_\nu$ changes as
\begin{align}\label{eq:app_d_rescale}
   |\vvec{\cc}_\nu| & \rightarrow |\vvec{\cc}_\nu| \alpha(\vvec{x} + \frac{1}{2} \vvec{\cc_\nu}^0)\times  \\ \nonumber & \qquad \qquad \qquad  \Bigg[  1 + \frac{s_1(\vvec{x})\left[(\vvec{\cc}_\nu^0 \cdot \hat{e}_1)^2 - (\vvec{\cc}_\nu^0 \cdot \hat{e}_2)^2 \right]}{2 |\vvec{\cc}_\nu^0|^2} 
    \\ & \qquad \qquad \qquad \qquad  \nonumber
      + \frac{s_2(\vvec{x})(\vvec{\cc}_\nu^0 \cdot \hat{e}_1) (\vvec{\cc}_\nu^0 \cdot \hat{e}_2)}{|\vvec{\cc}_\nu^0|^2} \Bigg] \, ,
\end{align}
where $\hat{e}_1, \hat{e}_2$ are the orthogonal basis vectors (x- and y- directions, respectively) in the reference space. The deformation also produces a reorientation of this local infinitesimal via
\begin{align}\label{eq:app_d_reorient}
   \mathrm{Arg}(\vvec{\cc}_\nu) \rightarrow \mathrm{Arg}(\vvec{\cc}_\nu) + \phi\left(\vvec{x} + \frac{1}{2} \vvec{\cc}_\nu\right) 
   \qquad \qquad \qquad  \\ \nonumber \qquad \qquad \qquad
   +
   \frac{s_2(\vvec{x})\left[(\vvec{\cc}_\nu \cdot \hat{e}_1)^2 - (\vvec{\cc}_\nu \cdot \hat{e}_2)^2 \right]}{2 |\vvec{\cc}_\nu|^2}    \\ \nonumber - \frac{s_1(\vvec{x})(\vvec{\cc}_\nu \cdot \hat{e}_1) (\vvec{\cc}_\nu \cdot \hat{e}_2)}{|\vvec{\cc}_\nu|^2} \, .
\end{align}
This information, combined with equation~(\ref{eq:orientation_fields}), allows us to write the angles of deviation
\begin{equation}\label{eq:cg_dev_angles}
\begin{aligned}
\delta \theta_A^{(\nu)}  = & \Delta T + \Delta \phi - \frac{\partial T}{\partial \alpha} \mathrm{Re}\left[ \cc_\nu \partial_z \alpha \right] - \mathrm{Re}\left[ \cc_\nu \partial_z \phi \right] \\
& - \frac{\mathrm{Re}[\cc_\nu^2]}{2|\cc_\nu|^2} \left[ \frac{\partial T}{\partial \alpha} \alpha s_1 + s_2 \right] + \frac{\mathrm{Im}[\cc_\nu^2]}{2|\cc_\nu|^2} \left[  s_1 - s_2 \frac{\partial T}{\partial \alpha} \alpha \right] \\
\delta \theta_B^{(\nu)}  = & \Delta T - \Delta \phi + \frac{\partial T}{\partial \alpha} \mathrm{Re}\left[ \cc_\nu \partial_z \alpha \right] - \mathrm{Re}\left[ \cc_\nu \partial_z \phi \right] \\
& + \frac{\mathrm{Re}[\cc_\nu^2]}{2|\cc_\nu|^2} \left[s_2 - \frac{\partial T}{\partial \alpha} \alpha s_1 \right] - \frac{\mathrm{Im}[\cc_\nu^2]}{2|\cc_\nu|^2} \left[  s_1 + s_2 \frac{\partial T}{\partial \alpha} \alpha \right] \\
\end{aligned}
\end{equation}
which will be input to determine the energy of hinge $\nu$. For convenience, we have moved to the complex formulation rather than the vector formulation where $z \equiv X_1 + i X_2$, and similar for vector quantities such as $\cc_\nu$. Our continuum energy density may now be constructed by summing the hinge energy equation~(\ref{eq:ang_param_energy}) over the four hinges in the unit cell via
\begin{equation}\label{eq:app_energy_sum}
   \Phi(z) = \frac{1}{A_{\text{cell}}}\sum_{\nu=1}^4 E_H^{(\nu)}
\end{equation}
where $A_{\text{cell}}=4 a^2 \mathrm{cos}(T_0)$ is the area of the unit cell in the reference space and $E_H^{(\nu)}$ is the energy from equation~(\ref{eq:ang_rescale_energy}) for hinge $\nu$. Inserting equations~\ref{eq:cg_dev_angles} and performing the sum yields
\begin{widetext}
\begin{equation}\label{eq:app_e_density_unsimplified}
\begin{aligned}
    \Phi(z) & = \frac{\ell^2}{a^2} \frac{g(\alpha)}{\mathrm{cos}(T_0)^2}  
    + \frac{(k_1(\alpha) + k_2(\alpha)) T'(\alpha)^2\alpha^2}{4 \mathrm{cos}(T_0)^2} s_1^2 
     + \frac{(k_1(\alpha) - k_2(\alpha))}{4 \mathrm{cos}(T_0)^2} s_2^2 \\ &
     + \left[ \frac{a^2 k_1(\alpha)}{4}(\frac{1}{\alpha^2} + T'(\alpha)^2)) + \frac{a^2 k_2(\alpha)}{4}(\frac{1}{\alpha^2} - T'(\alpha)^2)) \right] |\nabla \alpha|^2 + \frac{1}{\mathrm{cos}(T_0)} \left(   k_1(\alpha) + k_2(\alpha) \right) \left( \Delta T^2 + \Delta \phi^2 \right) \, .
     \end{aligned}
\end{equation}
\end{widetext}
The energy in equation~(\ref{eq:app_e_density_unsimplified}) includes all terms which are not higher order in either $\ell/a$ or the shears or $|\nabla \alpha|$. To reduce to an elasticity theory in terms of spatial deformation variables, we minimize equation~(\ref{eq:app_e_density_unsimplified}) with respect to the internal degrees of freedom $\Delta T$ and $\Delta \phi$. This leads to $\Delta T = 0$ and $\Delta \phi = 0$, which indicates that our mechanism-based guess for the square orientations in equation~(\ref{eq:orientation_fields}) was correct. The continuum elastic theory for the RS lattice is then
\begin{equation}\label{eq:app_elasticity_theory}
\begin{aligned}
    E =  \int \dee^2 \vvec{X} \frac{1}{2} \bigg\{ \frac{\ell^2}{a^2} M(\alpha) + a^2 \tilde{M}(\alpha)|\nabla \alpha|^2 +  \qquad \qquad \\ \qquad \qquad G_1(\alpha)s_1^2 + G_2(\alpha)s_2^2  \bigg\} \, .
    \end{aligned}
\end{equation}
Where the moduli
\begin{align}\label{eq:app_nonlinear_moduli}
    B(\alpha) & = \frac{2 g(\alpha)}{\mathrm{cos}(T_0)^2}, 
    \hspace{5cm}
    \\
    G_1(\alpha) & =  \frac{(k_1(\alpha) + k_2(\alpha)) T'(\alpha)^2\alpha^2}{2 \mathrm{cos}(T_0)^2}  
    , 
    \hspace{2cm}
    \\
    G_2(\alpha) & = \frac{(k_1(\alpha) - k_2(\alpha))}{2 \mathrm{cos}(T_0)^2}, 
    \hspace{5cm}
    \\
    \tilde{B}(\alpha) & = \frac{ k_1(\alpha)}{2}\left(\frac{1}{\alpha^2} + T'(\alpha)^2\right) 
    \hspace{5cm} \\ & \qquad \qquad \nonumber
    + \frac{ k_2(\alpha)}{2}\left(\frac{1}{\alpha^2} - T'(\alpha)^2\right) 
    ,
    \hspace{3cm}
\end{align}
capture the energy cost of dilation, pure shear, simple shear, and dilation gradients, respectively. Note that in the small hinge limit, the strain gradient modulus may be inferred from the shear moduli via the simple relation
\begin{equation}\label{eq:app_moduli_relate}
    \tilde{B}(\alpha) = \frac{1-\alpha^2 \mathrm{cos}^2(T_0)}{ \alpha^4} G_1(\alpha) + \frac{\mathrm{cos}^4(T_0)}{1-\alpha^2 \mathrm{cos}^2(T_0)} G_2(\alpha),
\end{equation}
where we have used the analogue for the mechanism in equation~(\ref{eq:real_mechanism}) to replace
\begin{equation}\label{eq:app_mech_relation}
    T'(\alpha) = \frac{-\mathrm{cos}(T_0)}{\sqrt{1 - \alpha^2 \mathrm{cos}^2(T_0)}} \, 
\end{equation}
excluding higher order terms proportional to $\ell / a$.


 \subsection{Symmetry-based construction of the energy functional}

 Here, we present an alternate derivation of an energy functional of the form equation~(\ref{eq:app_elasticity_theory}) without recourse to the exact RS microstructure, based instead upon its fundamental symmetries.
 
 We consider, then, a general elastic system that:
 
 \begin{itemize}
     \item has a dilational mechanism
     \item has standard elastic translational and rotational symmetries
     \item has a four-fold rotational symmetry
     \item has an $x\rightarrow -x$ mirror symmetry
     \item has a $y\rightarrow -y$ mirror symmetry (as implied by the previous two symmetries)
 \end{itemize}
 
 \noindent These last three properties arise from the \textit{p4g} wallpaper symmetry group, which applies to all states of the lattice along the uniform dilational mechanism.

 A uniform application of the mechanism will generate an energy density that we again label $(\ell/a)^2 M(\alpha)$, having identified the scaling with hinge thickness in \ref{app:hinges}. Additionally, energy terms may arise from other components of the deformation tensor: pure shear ($s_1$), simple shear ($s_2$) and rotation ($\phi$). However, rotation in particular can be easily eliminated: the system is invariant under a uniform rotation and equation~(\ref{eq:app_conf_closure}) (Cauchy-Riemann) shows that for dilation-dominated deformations, rotation gradients can be expressed in terms of dilation gradients instead (up to a small correction from the shears). Overall, our energy density can therefore be expressed in terms of three fields: $(s_1, s_2, \alpha)$ and gradients thereof. In general, this allows for terms proportional to $s_1^2, s_1 s_2, \partial_x \alpha \partial_y \alpha$ etc. Terms such as $\partial_x s_1 \partial_y s_1$ and $s_2 \partial_x^2 \alpha$ are also permitted, but these will be higher order in gradients and shears which are small due to the comparative size of the unit cell and design based on a mechanism, respectively. Note that, for present convenience, we are using notation different from the previous appendix, where $x\rightarrow X_1$ and $y \rightarrow X_2$.  
 
 Such energy terms are only permitted if they are invariant under the discrete symmetries of the undeformed or uniformly dilated lattice: a four-fold rotation about the center of a square and two mirror symmetries about the center of an open pore.
 
  \begin{table}[]
     \centering
     \begin{tabular}{c|c|c|c|c}
           & $s_1$ & $s_2$ & $\partial_x \alpha$ & $\partial_y \alpha$   \\ \hline
          $C_4$ & $-s_1$ & $-s_2$ & $-\partial_y \alpha$ & $\partial_x \alpha$ \\
          x-mirror & $s_1$ & $-s_2$ & $-\partial_x \alpha$ & $\partial_y \alpha$  \\
          y-mirror & $s_1$ & $-s_2$ & $\partial_x \alpha$ & $-\partial_y \alpha$
     \end{tabular}
     \caption{Action of symmetry operations on deformation fields}
     \label{tab:symmetries}
 \end{table}
 
 The action of the symmetry operations on the relevant fields are listed in Table.~\ref{tab:symmetries}. Enforcing that the energy remain the same before and after applying the x-mirror symmetry eliminates ($s_2$, $\partial_x \alpha$, $s_1 s_2$, $\partial_x \alpha \partial_y \alpha$, $s_2 \partial_y \alpha$, $s_1 \partial_x \alpha$ ) from the energy density. Similarly, the y-mirror symmetry eliminates the terms ($\partial_y \alpha$, $s_2 \partial_x \alpha$, $s_1 \partial_y \alpha$). Finally, the $C_4$ symmetry eliminates $S_1$, and further enforces that the energetic coefficient of the $(\partial_x\alpha)^2$ term match the coefficient of $(\partial_y\alpha)^2$. We do not consider terms involving second gradients of the dilation field, which are ruled out in the coarse-grained microscopic theory. 
 
 %
 %
 With this, the energy functional to lowest order in the small shears and dilation gradients must take the form
 \begin{equation}\label{eq:app_elasticity_theory2}
\begin{aligned}
    E = \int \dee^2 \vvec{x} \frac{1}{2} \bigg\{ \frac{\ell^2}{a^2} M(\alpha) + a^2 \tilde{M}(\alpha)|\nabla \alpha|^2 +  \qquad \qquad \\ \qquad \qquad G_1(\alpha)s_1^2 + G_2(\alpha)s_2^2  \bigg\} \, .
    \end{aligned}
\end{equation}
 %
 where the coefficients have been named to connect to the coarse grained theory in equation~(\ref{eq:app_elasticity_theory}).

 Importantly, while we have not invoked the specific nature of the RS mechanism or material properties to derive this elastic energy, is it specific to the point symmetry group of the lattice and a different energetic form is expected for other dilational metamaterials such as the kagome lattice.


\subsection{Stress of Near-Conformal Deformations of the RS Metamaterial}

Having constructed the Energy in equation~\ref{eq:app_elasticity_theory}, we would like to know the corresponding stress tensor, as this is the more common quantity used in nonlinear elasticity. However, this theory is both nonlinear and includes strain gradient terms, and we require a relation which appropriately incorporates these effects. To do this, we start with the essential information that the functional derivative with respect to displacement gives the force density as
\begin{equation}\label{eq:force_density1}
    f_i = -\frac{\delta E}{\delta u_i} \, .
\end{equation}
In addition, we know that the force density is the divergence of a stress tensor
\begin{equation}\label{eq:force_density2}
    f_i = \partial_j N_{ji} \, ,
\end{equation}
and so the process of deriving the stress is a matter of taking the functional derivative in equation~\ref{eq:force_density1} and fitting it into the form of equation~\ref{eq:force_density2}. Our energy functional is an integral over an energy density which is a function of our quantities $\alpha$ and $s_1$ and $s_2$. These are strain quantities, and may be expressed in terms of the Lagrangian strain
\begin{equation}\label{eq:Lagrange_strain}
    \varepsilon_{ij} = \frac{1}{2}(\partial_i u_j + \partial_j u_i + \partial_i u_k \partial_j u_k) = \frac{1}{2}(C_{ij} - \delta_{ij}) \, ,
\end{equation}
where $C_{ij} = F_{ki} F_{kj}$ is the right Cauchy-Green deformation tensor (i.e. the metric of deformation) and $F_{ij} = \partial_j u_i$ is the deformation gradient tensor. The variables in our energy functional equation~\ref{eq:app_elasticity_theory} may be written, to lowest order in $s_1, s_2$,  in terms of this strain via
\begin{align}\label{eq:app_strain_relations}
    \alpha & = \sqrt{\mathrm{det}[F]} = (\mathrm{det}[C])^{1/4} = (\mathrm{det}[2 \varepsilon + \mathbb{1}])^{1/4} \\
    s_1^2 & = \frac{1}{2} \frac{ \mathrm{Tr}\left[ C\cdot \sigma^{(3)} \right] }{\sqrt{\mathrm{det}\left[ C\right]}} =  \frac{ \mathrm{Tr}\left[ \varepsilon \cdot \sigma^{(3)} \right] }{\sqrt{\mathrm{det}\left[ 2 \varepsilon - \mathbb{1}\right]}} \\
    s_2^2 & = \frac{1}{2} \frac{ \mathrm{Tr}\left[ C\cdot \sigma^{(1)} \right] }{\sqrt{\mathrm{det}\left[ C\right]}} =  \frac{ \mathrm{Tr}\left[ \varepsilon \cdot \sigma^{(1)} \right] }{\sqrt{\mathrm{det}\left[ 2 \varepsilon + \mathbb{1}\right]}} \, .
\end{align}
Therefore, we should think of our energy functional as having the form
\begin{equation}
    E = \int \mathrm{d}^d x \Phi(\varepsilon_{ij}, \partial_k \varepsilon_{ij})
\end{equation}
With this, taking the functional derivative in equation~\ref{eq:force_density1} is a matter of using the chain rule for functional derivatives
\begin{equation}\label{eq:functional_chain}
    \frac{\delta E}{\delta u_l(x)} = \int \mathrm{d}^d x' \frac{\delta E}{\delta \varepsilon_{ij}(x')} \frac{\delta \varepsilon_{ij}(x')}{\delta u_l(x)} \, .
\end{equation}
Here, $\epsilon_{jk}$ may be thought of as a functional of $u_i$ and this derivative is taken using the definition of the functional derivative
\begin{equation}\label{eq:functional_deriv_general}
    \frac{\delta F[u]}{\delta u(x)} = \lim_{h\rightarrow 0} \frac{F[u(x')+ h \delta(x-x')] - F[u(x')]}{h}
\end{equation}
which becomes
\begin{align}\label{eq:functional_deriv_strain}
    \frac{\delta \varepsilon_{ij}[u(x')]}{\delta u_l(x)} = & \lim_{h\rightarrow 0} \frac{\varepsilon_{ij}[u(x')+ h \hat{e}_l\delta(x-x')] - \varepsilon_{ij}[u(x')]}{h} \\ 
    \nonumber & = \frac{\partial \epsilon_{ij}(x)}{\partial(\partial_k u_l(x))} \partial_k \delta(x-x') \, .
\end{align}
And using the form of equation~\ref{eq:Lagrange_strain}, we find
\begin{equation}\label{eq:functional_deriv_strain2}
    \frac{\delta \epsilon_{ij}[u(x')]}{\delta u_l(x)} =\frac{1}{2} \left( \delta_{lj} \partial_i + \delta_{il} \partial_j + \partial_i u_l \partial_j + \partial_j u_l \partial_i \right) \delta(x-x') \, .
\end{equation}
Plugging this back into equation~\ref{eq:functional_chain}, and simplifying a bit, we may find
\begin{equation}\label{eq:stress_relation}
N_{ij} = \frac{\delta E}{\delta \epsilon_{ik}} F_{kj} \, .
\end{equation}
where
\begin{equation}
    \frac{\delta E}{\delta \epsilon_{ij}} = \frac{\partial \Phi}{\partial \epsilon_{ij}} - \partial_k \frac{\partial \Phi}{\partial(\partial_k \epsilon_{ij})}
\end{equation}
is the standard functional derivative. As the divergence of this stress in terms of the reference space coordinates gives the force density in \emph{reference} space, this construction captures the Nominal stress (also known as Engineering stress and corresponding to the transpose of the first Piola-Kirchhoff stress) in accordance with the nonlinear elasticity literature~\cite{ogden1997non}. Using the standard conversion formulae taking us between the different standard nonlinear stress definitions, we find that the second Piola-Kirchhoff stress is given by the simple expression
\begin{equation}\label{eq:2pk_formula}
S_{ij} = \frac{\delta E}{\delta \epsilon_{ij}} \, .
\end{equation}

Inserting the actual energy from equation~\ref{eq:app_elasticity_theory}, we will find
\begin{align}\label{eq:2pk_final}
    S_{ij} = & \left[ T_1 \right] \delta_{ij} \\ \nonumber
    & - s_1 \left[ T_1 - \frac{G_1}{\alpha^2} - \frac{a^2}{\alpha s_1} \tilde{M} \partial_k \alpha \partial_k s_1 \right] \sigma_{ij}^{(3)}  \\ \nonumber
    & - s_2 \left[ T_1 - \frac{G_2}{\alpha^2} - \frac{a^2}{\alpha s_2} \tilde{M} \partial_k \alpha \partial_k s_2 \right] \sigma_{ij}^{(1)} \, ,
\end{align}
where
\begin{align}\label{eq:stress_special_terms}
    T_1 = & -\frac{a^2}{\alpha} \left[ \tilde{M} \partial_k \partial_k \alpha + \tilde{M}' \partial_k \alpha \partial_k \alpha  \right]  - 2 \frac{a^2}{\alpha^2} \tilde{M} \partial_k\alpha \partial_k\alpha \\ & \nonumber +  \frac{1}{2 \alpha} \Bigg[ \frac{\ell^2}{a^2} M' + a^2 \tilde{M} \partial_k\alpha \partial_k\alpha \\ & \nonumber \qquad \qquad+ \frac{s_1^2}{2}\left( G_1' - 4 \frac{G_1}{\alpha} \right) + \frac{s_2^2}{2}\left( G_2' - 4 \frac{G_2}{\alpha} \right)   \Bigg]  \, .
\end{align}
While the 2nd Piola-Kirchhoff doesn't admit a direct physical interpretation in terms of real traction forces across material surfaces, it is mathematically convenient and can easily be converted to the more physically relevant Cauchy stress with the formula
\begin{equation}
    \sigma_{ij} = \frac{1}{\text{det}[F]}F_{ik} S_{kl} F_{jl} \, .
\end{equation}
%


\subsection{Analytic prediction of deformation}



Having identified the effective continuum energy which will govern the deformation of the RS metamaterial, we would like to use it to predict material response to applied loads. Taking the limits of small hinge and small lattice spacing in equation~(\ref{eq:app_elasticity_theory}) leads to an energy which is comparatively very stiff against deformations that include shears, and the shear terms dominate the energy functional. In this limit, the material will choose to expel shear whenever possible, leading directly to the conformal maps (which definitionally exclude local shear) as the lowest energy deformations. With only shear terms in the energy, the conformal deformations form a highly degenerate space of ground states. The point displacements applied in the Finite Element simulations from \ref{app:op_derive} are not sufficient to fully break the degeneracy --- an infinite space of deformations satisfy the (point) boundary conditions without any shears, leaving no single prediction for the deformation. 

To break this degeneracy, we turn to perturbation theory, adding back in the dilation $M(\alpha)$ and dilation gradient $\tilde{M}(\alpha)$ terms. Predicting the deformation then becomes a search for the conformal map which minimizes the perturbative energy
\begin{equation}\label{eq:app_conformal_theory}
    E \sim \Delta E \equiv \int \dee \vvec{x} \frac{1}{2} \left\{ \frac{\ell}{a} M(\alpha) + a^2 \tilde{M}(\alpha) |\nabla \alpha|^2 \right\} \, .
\end{equation}
Shown in Fig.~\ref{fig:app_conformal_pred}-a, this reduced form is a reasonable approximation to the true energy, with the shear energy roughly an order of magnitude smaller at small hinge size. This energy gap will be even more significant, and the perturbative approximation better, for larger material samples with more unit cells which are closer to the continuum limit. As described in the main text, this process of obtaining and using this effective theory constitutes our \textbf{conformal elasticity}. 


%
%


\subsubsection{Analytic prediction of linear deformation using effective conformal theory \label{app:linear_predictions}}


\begin{figure*}[!t]  
\begin{center}
    \includegraphics[width=0.9\textwidth]{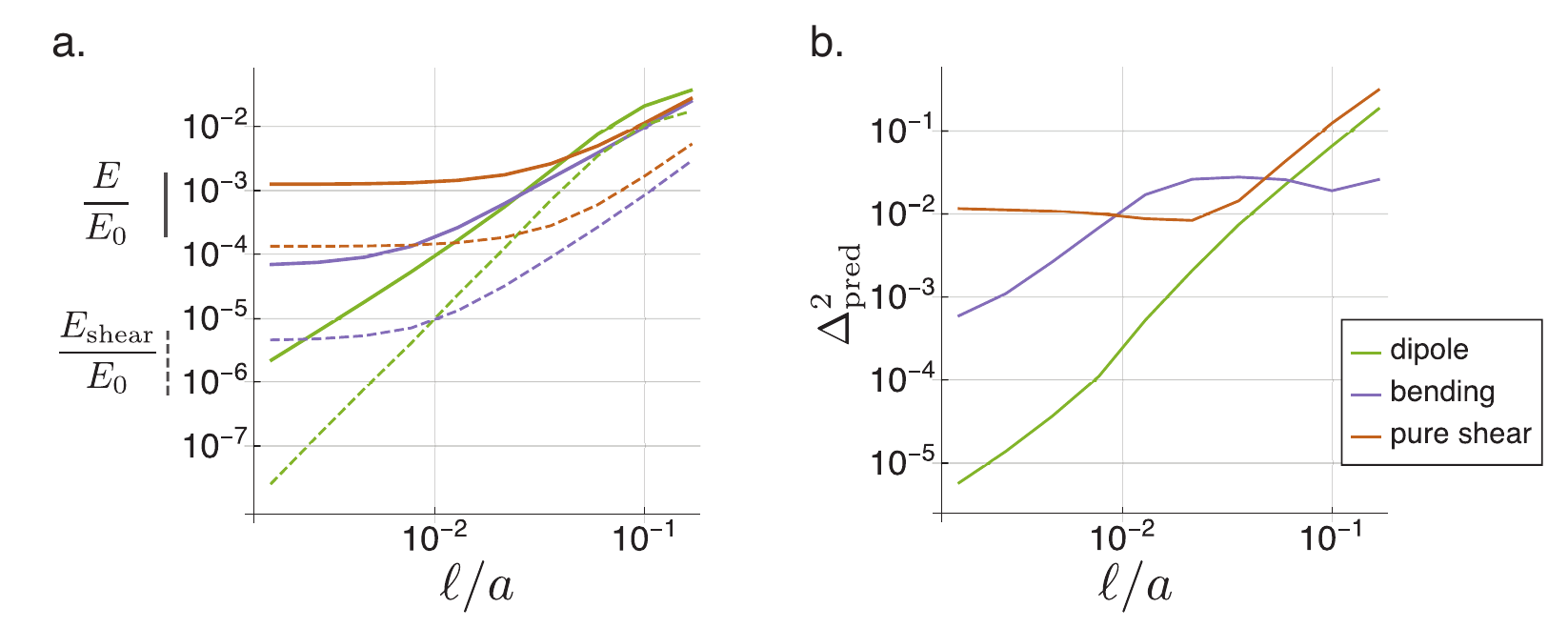}
    \caption{\label{fig:app_conformal_pred} 
     (\textbf{a}) We show that going from equation~(\ref{eq:app_elasticity_theory}) to equation~(\ref{eq:app_conformal_theory}) is a good approximation by showing that the contribution to the energy from the shear terms is an order of magnitude smaller (or more) in the limit of small hinges. Energies are normalized by $E_0$, which is the amount of energy required to create the same coarse strain magnitude in the absence of the pore structure, and which showcases also the comparative softness of the porous metamaterial. The energy in equation~(\ref{eq:app_conformal_theory}) may also be used to generate predictions of small deformations which have very small fractional mean squared error shown in (b).
 }
    \end{center}
\end{figure*}

The energy in equation~(\ref{eq:app_conformal_theory}) is highly nonlinear and minimization requires prior knowledge of the nonlinear stiffnesses $k_1(\alpha), k_2(\alpha), g(\alpha)$, which are not easily obtained. Even with knowledge of these moduli, analytic solutions to such nonlinear mechanics problems are hard to come by. However, taking the theory to lowest order in strains and rewriting in terms of a complex displacement $u(z) = f(z) - z = u_x(z) + i u_y(z)$ yields
\begin{equation}\label{eq:app_e_linearized}
    E = \int \dee z \dee \bar{z} \left\{ K \mathrm{Re}(\partial_z u)^2 + \frac{1}{2}\tilde{K} |\partial^2_z u|^2 \right\} + E_{constr}
\end{equation}
where $\partial_z = 0.5(\partial_x - i \partial_y)$ is a standard complex derivative~\cite{england2003complex} and 
\begin{equation}\label{eq:app_linear_bulk_moduli}
    \begin{aligned}
    K &= \frac{\ell^2 g''(1)}{4 a^2 \mathrm{cos}(T_0)^2} \\
    \tilde{K} &= a^2 \tilde{M}(\alpha=1)  \, 
    \end{aligned}
\end{equation}
%
%
are the bulk and bulk gradient moduli, respectively. Note the unknown function $g$, defined in \ref{app:hinges}, which captures the  mechanism energy. In order to incorporate the applied point motions (constraints), Lagrange multiplier terms have been added to the energy functional as
\begin{equation}\label{eq:app_point_motions}
    E_{constr} = \int \dee z \dee \bar{z} \left\{ \sum_k \lambda_k (u_k - u(z_k)) \right\} \, ,
\end{equation}
where $\{ u_k\}$ are the displacements prescribed at locations $\{ z_k \}$. 

The energy in equation~(\ref{eq:app_e_linearized}) will be used to select between conformal displacements $u$ which may be written as an expansion $u(z) = \sum_n C_n z^n$. As higher coefficients $C_n$ generate sharper and sharper features, we can cut off the expansion and then solving the equations of energy minimization is reduced to a linear algebra problem. 

The above is not yet sufficient to generate predictions, as the moduli $K$ and $\tilde{K}$ remain undetermined. Conveniently, the coarse-graining procedure from \ref{app:coarse_grain} offers a relation between these moduli which is exact in the limit of small hinges
\begin{equation}\label{eq:moduli_related}
  \tilde{K} = a^2\left( \mathrm{sin}(T_0)^2 G_1(\alpha=1) + \frac{\mathrm{cos}(T_0)^2}{\mathrm{tan}(T_0)^2} G_2(\alpha=1)  \right) \, .
\end{equation}
With this, the strain level moduli $(K, G_1, G_2)$ obtained numerically in the methods section of the main text constitute all necessary information to make linear deformation predictions. For simulations with hinge thickness $\ell_0=0.1$mm these moduli are $(K, \tilde{K})  = (35\text{Pa*m}, 125 \text{Pa*m}^3)$. To find predictions at different hinge thicknesses $\ell$, we rely on the coarse graining results, which indicate that $K \sim \ell^2$ while all other moduli $G_1, G_2, \tilde{K}$ are independent of $\ell$. Therefore, we write $K(\ell) = \frac{\ell^2}{\ell_0^2} K(\ell_0) $ in terms of the known $K(\ell_0)$ and use these values to solve the linear algebra problem for the coefficients $\{ C_n\}$ and therefore our prediction for $u(z)$. These predictions capture all but a small portion  $\Delta_{pred}^2$ of the error, as displayed in Fig.~\ref{fig:app_conformal_pred}-b. This completes

\subsection{Lagrange multiplier Approach to Conformal Elasticity}

In this Appendix, we employ a more conventional method of enforcing the shear-free conditions on the RS metamaterial by the introduction of Lagrange multipliers $\lambda_1, \lambda_2$. This produces insight into the shear stresses enforcing this shear-free condition. For simplicity, we consider a system without gradient terms, so that the energy takes the form

\begin{align}
    E = \int d^2 \mathbf{R} \, \left[ b(J) + \lambda_1 (F_{11} - F_{22}) + \lambda_2(F_{12} + F_{21}) \right],
\end{align}

\noindent where $F_{ij}$ is the deformation tensor, $J = \alpha^2$ is the determinant of its matrix form and the general dependence on the reference coordinate has been suppressed.

The conditions for equilibrium are that the energy cannot be lowered by any movement of the material from its target position, $\mathbf{r}(\mathbf{R})$. This can be expressed as a functional derivative (See, e.g., Lazar and Kirchner~\cite{Lazar2007} or Saremi and Rocklin~\cite{Saremi2020}):
\begin{align}
    \frac{\delta E}{\delta r_1(\mathbf{R'})} = \frac{\delta E}{\delta r_2(\mathbf{R'})}=0.
\end{align}
These functional derivatives inherit many properties of conventional derivatives and in particular if the energy depends on the position only through an intermediate function, such as the deformation tensor, we have the chain rule:
\begin{align}
    \frac{\delta E}{\delta r_i(\mathbf{R'})} = \int d^2 \mathbf{R}'' \, \frac{\delta E}{\delta F_{jk}(\mathbf{R''})}\frac{\delta F_{jk}(\mathbf{R''})}{\mathbf{r}(\mathbf{R}')}.
\end{align}
Consequently, after some algebra, we obtain our equilibrium conditions:
\begin{align}
    b''(J)\left[F_{11} \partial_1 J + F_{12} \partial_2 J \right]+\partial_1 \lambda_1 + \partial_2 \lambda_2 = 0  ,  \\
    b''(J)\left[F_{11} \partial_2 J - F_{12} \partial_1 J \right]-\partial_2 \lambda_1 + \partial_1 \lambda_2 = 0  .
\end{align}
\noindent In obtaining this expression, we have already used that $F_{22} = F_{11}, F_{21} = - F_{12}$ (the constraints). To this, we now add our compatibility condition, as discussed in the Methods section of the main text. We find that this compatibility condition is equivalent to

\begin{equation}
  (\partial_1 + i \partial_2)(F_{11} - i F_{12}) = 0,
\end{equation}
or
\begin{equation}
  \partial_{\bar{z}} f' = 0
\end{equation}
where
\begin{align}
  z \equiv x+ i y,\\
  \partial_{\bar{z}} = \frac{1}{2}(\partial_x + i \partial_y)\\
  f' \equiv F_{11} - i F_{12}.
\end{align}

We now note that $J = |f'|^2$ and we can take our two original real equilibrium conditions and add the first equation to the second equation multiplied by $i$, and obtain the equivalent complex condition
\begin{align}
    b''(J) |f'^2| \bar{f''} + \partial_z \lambda=0,\\
    \lambda(z,\bar{z}) \equiv \lambda_1 + i \lambda_2.
\end{align}
governing the constrained equilibrium states. Thus, there are stresses that are supported even in the absence of deformation ($f'(z)=0$): the structure supports shear stresses that are functions of $\bar{z}$ only. That is, they are antiholomorphic, whereas the low-energy deformations are holomorphic, or conformal. In addition, there is an additional shear stress proportional to $b''(J)$ that forms when there is a nonzero bulk modulus and a spatially varying deformation tensor. Similar results arise more generally when dilation gradient terms are included, still allowing for an undetermined antiholomorphic space of shear stresses.


\subsection{Boundary method for analytic prediction of conformal deformation \label{app:boundary_method}}



As described in the main text, materials in the conformal limit admit a novel boundary method for analytic control of deformation. Here we give more explicit analytic recipes both for continuous analytic description of the  metamaterial and for a discrete approximation. 

As noted in the main text, a planar conformal deformation may be described by a complex analytic function $z \rightarrow f(z)$. The function $f(z)$ obeys $\partial_{\bar{z}}f = 0$ ($\bar{z}$ is the complex conjugate of $z$), and as a result admits well defined $z$-derivatives everywhere in its domain and can be expanded in a series as in equation~(\ref{eq:conformal_expand_app}). These analytic properties are preserved when the derivative is taken $f' \equiv \partial_z f$ and when we take the logarithm, defining $g \equiv \mathrm{ln}(f')$. Note that the logarithm has the potential to introduce nonanalyticity, yet avoiding nonphysical material configurations such as inversion and collapse into a point avoids any such problems. The function $f'$ may be written in modulus-argument form as
\begin{equation}
    f'(z) = \alpha(z,\bar{z}) \mathrm{exp}(i \phi(z, \bar{z})) \, .
\end{equation}
In finite strain theory, $f'$ captures the deformation tensor, the function $\alpha$ will capture the local isotropic rescaling of area $\mathrm{d}A \rightarrow \alpha^2 \mathrm{d}A$ due to deformation, and $\phi$ captures the coarse material rotation (e.g. reorientation of the entire unit cell). Taking the logarithm, our function $g$ may be written
\begin{equation}
    g = \mathrm{ln}(\alpha) +  i \phi \, .
\end{equation}
The real part of this function now depends only on the scaling factor $\alpha$ while the imaginary part depends only on the rotation $\phi$. It is well known that the real and imaginary parts of a complex analytic function each separately satisfy the Laplace equation, which appears as $4 \partial_z \partial_{\bar{z}} \mathrm{ln}(\alpha) = 0$ in complex form for the real part of $g$. The Laplace equation admits a unique solution given Dirichlet data (the local value of $\mathrm{ln}(\alpha)$) all along a closed boundary. Solving this for the real part of $g$, the boundary data for $\mathrm{ln}(\alpha)$ comes from local lattice dilations along the boundary. Therefore, given prescribed dilation data all along the boundary of a closed material domain, the dilations throughout the bulk are determined by the Laplace equation.

Next we note that full knowledge of $\mathrm{Re}[g]$ across the domain may be used to infer the full analytic function $g$ up to a purely imaginary constant. Recipes to infer $g$ such as the method of Milne-Thompson~\cite{Milne-Thomson1937} or that applied by John d'Angelo in Ref.~\cite{stone2009mathematics} rely on the Cauchy-Riemann equations. Employing the method of John d'Angelo we may write the full function
\begin{equation}
    g(z) = 2 \, \mathrm{ln}\left(\alpha\left( x \rightarrow \frac{z}{2} \, , \, y \rightarrow \frac{z}{2 i}\right)\right) + i \phi_0
\end{equation}
in terms of the internal dilation field $\alpha(x,y)$ up to an undetermined material rotation $\phi_0$. Finally, taking the exponential $f' = \mathrm{exp}(g)$ yields a complete prediction of the deformation tensor throughout the bulk. Integrating $f'$ along a path from $z_i$ yields a prediction for the relative final position of the material point initially located at $z_f$
\begin{equation}
    f(z_f) - f(z_i)  = \int_{z_i}^{z_f} \mathrm{exp}\left[ g  \right] \mathrm{d}z \, .
\end{equation}
Importantly, the conformal property guarantees that this integral is not path dependent. 

In summary, the full recipe for inferring the spatial mapping $f(z)$ from the boundary is as follows:
\begin{itemize}
  \item solve for $\alpha(x,y)$ using the boundary conditions and the equation $\nabla^2 \ln (\alpha) = 0$
  \item insert to determine the function $f'(z) = \exp \left[ 2 \ln(\alpha(\frac{z}{2}, \frac{z}{2 i}))  + \phi_0 i \right] $ where $\phi_0$ is the undetermined global rotation.
  \item integrate to find $f(z)$
\end{itemize}
We remark that the analytic viability of this recipe depends on the ability to generate solutions to the Laplace equation on the chosen domain, and to integrate them. The examples of square and circular domains provide convenient solvable examples. However, the true strength of our method lies in the reverse application. Given a desired conformal deformation $f(z)$, this may be prescribed by simply taking the derivative and evaluating the magnitude at the boundary, yielding the appropriate boundary dilation pattern to actuate. It is the uniqueness of the inverse recipe shown here (up to overall translations and rotations) which guarantees only the target map can satisfy these boundary conditions. 

Finally, the particular choice of mechanism sets limitations to the amount the metamaterial can be dilated and contracted to avoid self intersection and material over-extension. For instance, the RS mechanism requires $\alpha$ to be everywhere confined to the range $1/\sqrt{2} < \alpha < 1$, when starting from the fully dilated configuration. However, very conveniently, it turns out that any sufficiently smooth choice of boundary dilations which do not over- or under-dilate the RS lattice on the boundary, will also generate a conformal deformation in the interior which does not go beyond the valid dilation range of the specific mechanism. This is due to the maximum modulus principle which states that both the maximum and minimum dilations of a conformal map will occur at the boundary.

\subsubsection{Discrete Inference Method}
For most material domain shapes and boundary dilation patterns, a closed-form solution for $f(z)$ is not guaranteed. We therefore illustrate an alternative scheme that allows for quick and accurate inference of a map from discrete boundary data, and which generates the predictions in main text Fig.~4.

In this method, we first identify the boundary via a set of discrete points $\{z_k\}$ and the corresponding local dilations $\{\alpha_k\}$. Points should be densely spaced compared to the lengthscale of variation of $\alpha_k$. And, naturally, they should approximately trace a single continuous path enclosing a simply-connected region of the plane. Following this, we choose a cutoff $N$ in the number of coefficients of
\begin{equation} \label{eq:app_g}
g(z) = \sum_{n=0}^{N-1} C_n z^n.
\end{equation}
$N$ should be significantly less than half the number of boundary points $M$ in order to avoid overfitting.

Next we demand that our function $g(z)$ satisfy the boundary conditions by minimizing the error
\begin{equation}\label{eq:app_err}
    \textrm{err} = \sum_k^M (\mathrm{Re}[g(z_k)] - \ln(\alpha_k) )^2 \, .
\end{equation}
Inserting equation~(\ref{eq:app_g}), we minimize this error with respect to the coefficients $C_n = A_n + i B_n$ yielding the equations
\begin{equation}\label{eq:app_min_g}
    \begin{aligned}
    0 = & \frac{\partial [\textrm{err}]}{\partial A_l} = f^{(A)}_l + \sum_n^N A_n Q_{ln} + \sum_n^N B_n R_{ln} \\
    0 = & \frac{\partial [\textrm{err}]}{\partial B_l} = f^{(B)}_l + \sum_n^N A_n S_{ln} + \sum_n^N B_n T_{ln} \, ,
    \end{aligned}
\end{equation}
where
\begin{equation}\label{eq:app_min_g_mats}
    \begin{aligned}
    Q_{ln} = & \sum_k^M \left( r_k^{n+l} \mathrm{cos}(n\theta_k) \mathrm{cos}(l\theta_k) \right) \\
    R_{ln} = & -\sum_k^M \left( r_k^{n+l} \mathrm{sin}(n\theta_k) \mathrm{cos}(l\theta_k) \right) \\
    S_{ln} = & -\sum_k^M \left( r_k^{n+l} \mathrm{cos}(n\theta_k) \mathrm{sin}(l\theta_k) \right) \\
    T_{ln} = & \sum_k^M \left( r_k^{n+l} \mathrm{sin}(n\theta_k) \mathrm{sin}(l\theta_k) \right) \\
    f^{(A)}_{l} = & -\sum_k^M \left( \mathrm{ln}(\alpha_k) r_k^{l} \mathrm{cos}(l\theta_k) \right) \\
    f^{(B)}_{l} = & \sum_k^M \left( \mathrm{ln}(\alpha_k) r_k^{l} \mathrm{sin}(l\theta_k) \right)  \, , \\
    \end{aligned}
\end{equation}
and we have expressed $z_k = r_k e^{i\theta_k}$ in a complex polar form. equation~(\ref{eq:app_min_g}) reduces the inference of $g$ to a linear algebra problem which may be readily solved using built-in tools in, e.g., Mathematica. Note importantly that the row and column corresponding to $B_0 = \phi_0$ (the undetermined global rotation) are zero throughout. This row-column pair is, in general,  the only one that needs to be removed before numerically solving. 

The above yields the coefficients $C_n$ of $g(z)$, which is inserted into the exponential function to yield a prediction for $\partial_z f$ and therefore constitutes a full prediction of the deformation tensor everywhere inside the domain. In order to generate the displacement predictions shown in main text Fig.~3,  we employ the built-in numerical integrators in Mathematica.


\subsection{Boundary Control Simulation Protocol}

\begin{figure*}[!t]  
\begin{center}
    \includegraphics[width=0.6\textwidth]{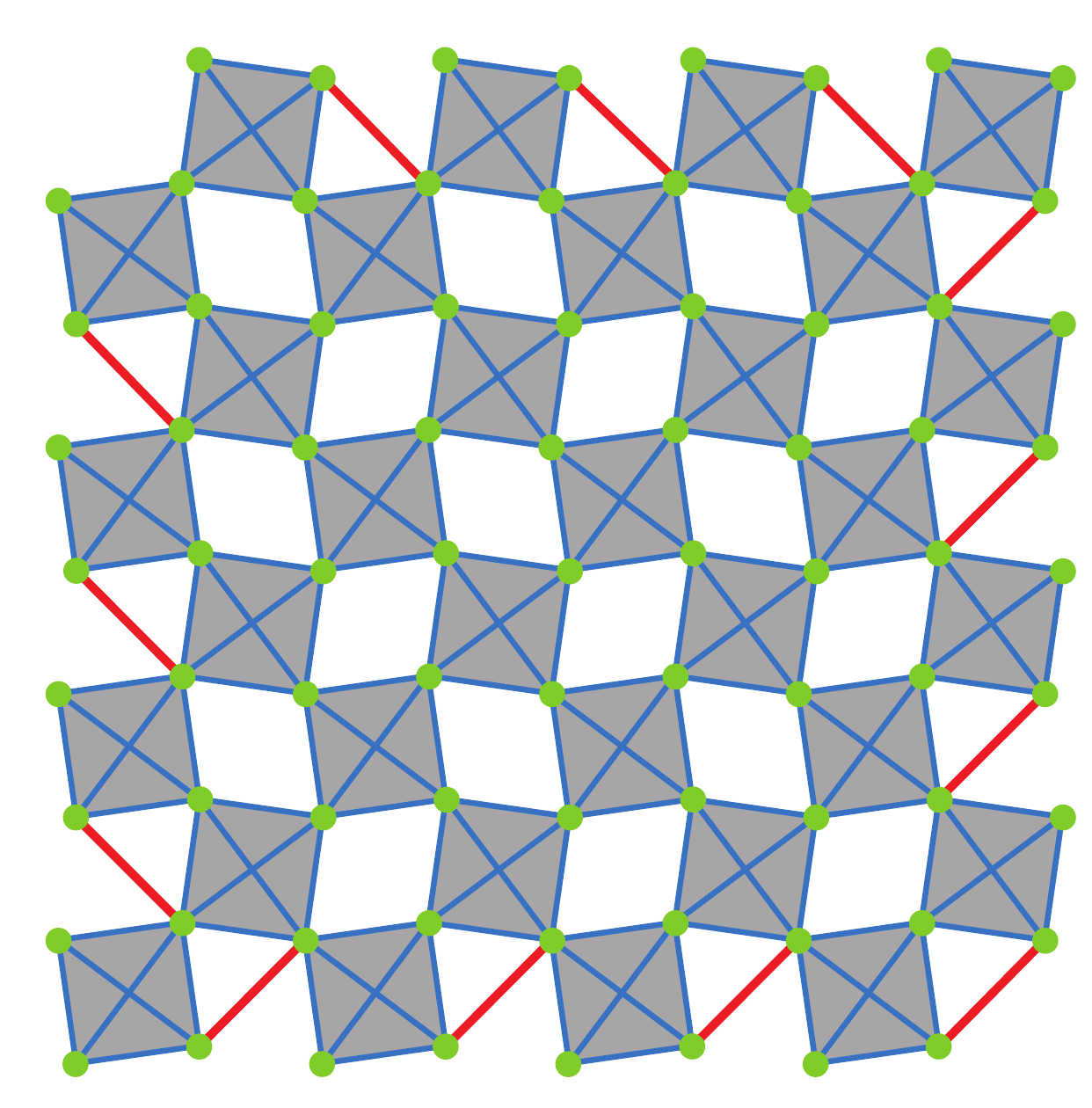}
    \caption{\label{fig:app_spring_model}  A simple spring model of RS metamaterial and boundary control. The numerical method employed to investigate the viability of boundary control of the RS elastic structure is accomplished using a structure of linear springs (blue lines) connect at frictionless nodes (green dots), approximating a collection of rigid squares (grey regions, displayed for visual convenience). Additional springs are added at the boundary (red lines) and the rest lengths varied to reliably actuate soft modes as shown in the main text Fig.4.
 }
    \end{center}
\end{figure*}

To probe the viability of boundary control of soft conformal modes, we perform numerical simulations of a simplified version of the RS lattice. In this case the lattice is composed purely of Hookean springs of identical stiffness $k$ connected at frictionless nodes. As shown in Fig.~\ref{fig:app_spring_model}, each rigid square element (grey) is emulated by a grouping of six springs. While a single cross-spring is sufficient to render a square element rigid, the inclusion of both ensures that the spring ensemble will obey the same symmetry properties as the elastic RS metamaterial. 

To actuate a particular soft conformal mode, stiff springs are added at the boundary, as shown in Fig.~\ref{fig:app_spring_model}. These boundary springs are taken to have stiffness $10^4 k$, so that they will act as rigid constraints, realizing their rest lengths quite accurately compared to the soft mechanics of the bulk material. While the rest lengths in the interior ( blue bonds) are chosen to leave the square shape at zero energy, the rest lengths of the boundary springs are varied to match some target local dilation. For three patterns of boundary dilation, we set these rest lengths and identify force balanced states using the conjugate gradient numerical minimization procedure ``minimize'' from the scipy.optimize toolkit in python. At the same time, using the methods from \ref{app:boundary_method}, and as illustrated in the main text Fig.3d\&e, predictions for displacements may be generated from this input set of boundary dilations. As shown in main text Fig.4, these are in good qualitative agreement with the numerically identified force balanced configurations.

\pagebreak

\bibliography{bibliofile}
\bibliographystyle{naturemag}

\end{document}